\documentclass[acmsmall]{acmart}

\AtBeginDocument{%
  }

\usepackage{multirow}
\usepackage{booktabs}
\usepackage{xcolor}
\usepackage{graphicx}
\begin{document}

\title{Generative Adversarial Networks for Image Super-Resolution: A Survey}


\author{Ziang Wu}
\affiliation{%
  \institution{School of Engineering, The Hong Kong University of Science and Technology}
  \streetaddress{Clear Water Bay, Kowloon}
  \city{Hong Kong}
  \country{China}
}
 \email{ziang.wu@connect.ust.hk}

\author{Xuanyu Zhang}
\affiliation{%
  \institution{School of Software, Northwestern Polytechnical University}
  \streetaddress{1 Dongxiang Rd}
  \city{Xi’an}
  \state{Shaanxi}
  \country{China}
  \postcode{710129}
}
  \email{xuanyuzhang@mail.nwpu.edu.cn}

\author{Yinbo Yu}
\affiliation{%
  \institution{College of Artificial Intelligence, Nanjing University of Aeronautics and Astronautics}
  \city{Nanjing}
  \state{Jiangsu}
  \country{China}
  \postcode{211106}
}
 \email{yinboyu@nuaa.edu.cn}

\author{Qi Zhu}
\affiliation{%
  \institution{College of Artificial Intelligence, Nanjing University of Aeronautics and Astronautics}
  \city{Nanjing}
  \state{Jiangsu}
  \country{China}
  \postcode{211106}
}
   \email{zhuqi@nuaa.edu.cn}

\author{Jerry Chun-Wei Lin}
\affiliation{%
  \institution{Department of Distributed Systems and IT Devices, Silesian University of Technology}
  \city{Gliwice}
  \country{Poland}
}
\email{jerrylin@ieee.org}

\author{Chunwei Tian\authornotemark}
\affiliation{%
  \institution{School of Computer Science and Technology, Harbin Institute of Technology}
  \streetaddress{92 Xidazhi Street}
  \city{Harbin}
  \state{Heilongjiang}
  \country{China}
  \postcode{150001}
}
\email{chunweitian@hit.edu.cn}
\authornote{Corresponding Author: Chunwei Tian. Email: chunweitian@hit.edu.cn}
\renewcommand{\shortauthors}{Wu et al.}

\begin{abstract}
Single image super-resolution (SISR) has played an important role in the field of image processing. Recent generative adversarial networks (GANs) can achieve excellent results on low-resolution images. However, there are little literatures summarizing different GANs in SISR. In this paper, we conduct a comparative study of GANs from different perspectives. We begin by surveying the development of GANs and popular GAN variants for image-related applications, and then analyze motivations, implementations and differences of GANs based optimization methods and discriminative learning for image super-resolution in terms of supervised, semi-supervised and unsupervised manners, where these GANs are analyzed via integrating different network architectures, prior knowledge, loss functions and multiple tasks. Secondly, we compare \textcolor{black}{the performances} of these popular GANs on public datasets via quantitative and qualitative analysis in SISR. Finally, we highlight challenges of GANs and potential research points for SISR.
\end{abstract}

\begin{CCSXML}
<ccs2012>
   <concept>
       <concept_id>10002944.10011122.10002945</concept_id>
       <concept_desc>General and reference~Surveys and overviews</concept_desc>
       <concept_significance>500</concept_significance>
       </concept>
 </ccs2012>
\end{CCSXML}

\ccsdesc[500]{General and reference~Surveys and overviews}

\keywords{SISR, GANs, small samples, optimization methods and discriminative learning}

\maketitle

\section{Introduction}
Single image super-resolution (SISR) is an important branch in the field of image processing \cite{wang2020deep}. \textcolor{black}{It aims} to recover a high-resolution (HR) image over a low-resolution (LR) image \cite{yang2019deep}, \textcolor{black}{leading to its} wide applications in medical diagnosis \cite{isaac2015super}, video surveillance \cite{zhang2010super} and disaster relief \cite{zha2015bayesian} etc. For instance, in the medical field, obtaining higher-quality images can help doctors accurately detect diseases \cite{huang2017simultaneous}. Thus, studying SISR is very meaningful to academia and industry.

To address SISR problem, researchers have developed a variety of methods based on degradation models of low-level vision tasks \cite{zhou2014local}. There are three categories for SISR in general, i.e., image itself information, prior knowledge and machine learning. In the image itself information, directly amplifying resolutions of all pixels in a LR image through an interpolation way to obtain a HR image was a simple and efficient method in SISR \cite{park2003super}, i.e., nearest neighbor interpolation \cite{rukundo2012nearest}, bilinear interpolation \cite{li2001new} and bicubic interpolation \cite{keys1981cubic}, etc. It is noted that in these interpolation methods, high-frequency information is lost in the up-sampling process \cite{park2003super}, which may decrease performance in image super-resolution. Alternatively, reconstruction-based methods were developed for SISR, \textcolor{black}{according to optimization methods} \cite{hardeep2013survey}. That is, mapping a projection into a convex set to estimate the registration parameters can restore more details of SISR \cite{stark1989high}. Although the mentioned methods can overcome the drawbacks of image itself information methods, they still suffered \textcolor{black}{from} the following challenges: non-unique solution, slow convergence speed and higher computational costs. To prevent this phenomenon, the \textcolor{black}{prior} knowledge and image itself information were integrated into a frame to find an optimal solution to improve the quality of the predicted SR images \cite{elad1997restoration,irani1991improving}. Besides, machine learning methods can be presented to deal with SISR, according to relation of data distribution \cite{sun2008image}. There are also many other SR methods \cite{sun2008image,yan2015single} that often adopt sophisticated prior knowledge to restrict the possible solution space with \textcolor{black}{the} advantage of generating flexible and sharp detail. However, the performance of these methods rapidly degrades  when the scale factor is increased, and these methods tend to be time-consuming \cite{nasrollahi2014super}.

\textcolor{black}{To obtain a better and more efficient SR model, a variety of deep learning methods were applied to a large-scale image dataset to solve the super-resolution tasks \cite{guo2023pft,gao2023ctcnet}.} For instance, Dong et al. proposed a super-resolution convolutional neural network (SRCNN) based pixel mapping that used only three layers to obtain stronger learning ability than these of some popular machine learning methods on image super-resolution \cite{dong2015image}. Although the SRCNN had a good SR effect, it still faced problems in terms of shallow architecture and high complexity. To overcome challenges of shallow architectures, Kim et al. \cite{kim2016accurate} designed a deep architecture by stacking some small convolutions to improve performance of image super-resolution. Tai et al. \cite{tai2017image} relied on recursive and residual operations in a deep network to enhance learning ability of a SR model. To further improve the SR effect, Lee et al. \cite{lim2017enhanced} used weights to adjust residual blocks to achieve better SR performance. To extract robust information, the combination of traditional machine learning methods and deep networks can restore more detailed information for SISR \cite{wang2015deep}. For instance, Wang et al. \cite{wang2015deep} embedded sparse coding method into a deep neural network to make a tradeoff between performance and efficiency in SISR. To reduce the complexity, an up-sampling operation is used in a deep layer in a deep CNN to increase the resolution of low-frequency features and produce high-quality images \cite{dong2016accelerating}. For example, Dong et al. \cite{dong2016accelerating} directly exploited the given low-resolution images to train a SR model for improving training efficiency, where the SR network used a deconvolution layer to reconstruct HR images. There are also other effective SR methods. For example, Lai et al. \cite{lai2017deep} used Laplacian pyramid technique into a deep network in shared parameters to accelerate the training speed for SISR. Zhang et al. \cite{zhang2018image} guided a CNN by attention mechanisms to extract salient features for improving the performance and visual effects in image SISR. \textcolor{black}{Dynamic super resolution network (DSRNet) \cite{tian2024imagesuperresolutiondynamicnetwork} proposes a dynamic network architecture that utilizes a dynamic gate mechanism to dynamically adjust network parameters to adapt to different scenarios, significantly enhancing the robustness and applicability of the image super-resolution model in complex scenes. Tian et al. \cite{tiancosine} proposed a cosine network-based method for SISR, which introduced odd and even enhancement blocks to extract complementary homologous structural information and incorporated a cosine annealing mechanism to optimize the training process, thereby achieving superior performance on multiple public datasets.} 

Although mentioned SR methods have obtained excellent effects in SISR with certain scales, they may suffer from poor adaptability for real scenes. To address this problem, generative methods, i.e., flow-based models \cite{lugmayr2020srflow}, VAEs \cite{heydari2020srvae}, diffusion model \cite{li2022srdiff} are developed. 
%张璇昱改
Specifically, flow-based models directly calculate the mapping relationship between the normal distribution and the target distribution function to generate the target data \cite{jo2021srflow}. This processing brings both reversible data and huge computation cost. Variational Autoencoders (VAEs) extract hidden variable distributions through encoders, and then randomly sample them to generate new images \cite{chira2022image}. The training process of VAEs is stable, but the quality of generated images limits its application. Diffusion models generates images with outstanding diversity via iterating denoising \cite{gao2023implicit}. However, it causes that the inference processing takes more time.

With the development of hardware devices, varieties of captured images are more\cite{li2025evasr}. However, scenes in the real world are varying, the number of obtained images are insufficient for real applications.  To address this problem, generative methods, i.e., flow-based models, VAEs, diffusion model are developed. 
Specifically, flow-based models directly calculate the mapping relationship between the normal distribution and the target distribution function to better generate similar samples as given samples \cite{jo2021srflow}. Although it has better generative effect, it suffers from a challenge of huge computational cost. To solve unstable training of GANs\cite{gan2025generative}, variational autoencoders (VAEs) used encoders to extract hidden variable distributions to overcome wave of training process \cite{chira2022image}. Although VAEs can be useful to accelerate training, they are limited by blurry image super-resolution.  To enhance adaptive ability of obtained super-resolution models, diffusion models can use game strategies between adding and reducing noise to learn an unsupervised super-resolution model. Although a diffusion model can enhance robustness of obtained image super-resolution model, it may cause bigger computational costs and consume more resources. Taking into mentioned analysis account, GANs based game strategies are good tools of generative methods for image super-resolution.
%$三种方法实现原理，作用，优缺点$ 基于流的模型直接计算从标准分布到目标分布函数之间的映射关系来生成目标数据。优点是数据转换通常是可逆的，缺点是计算量太大。VAEs通过编码器从已有的图像中提取隐变量分布，再从隐变量中随机采样生成图像.VAEs的训练过程很稳定，但是生成图像的质量限制了它的应用。扩散模型通过迭代去噪生成具有多样性的图像。(优点就是生成图像的多样性)然而，这会导致推理处理花费更多的时间。
% 
\begin{figure}[h!]
  \centering
  \includegraphics[width=0.6\linewidth]{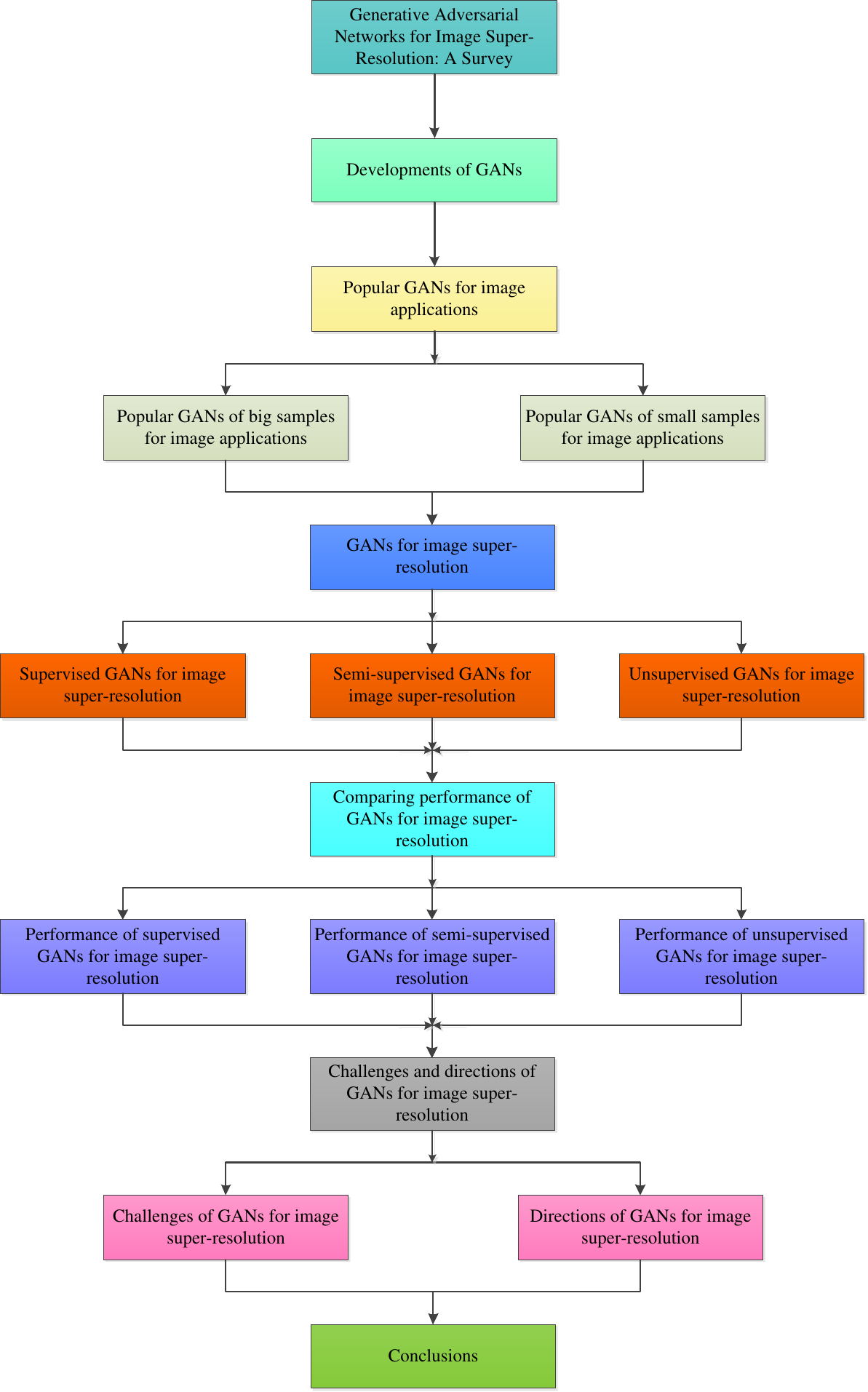}
  \caption{\textcolor{black}{Outline of this overview. It mainly consists of basic frameworks, categories (i.e., supervised, semi-supervised and unsupervised GANs), performance comparison, challenges and potential directions.}}
  \label{fig:1}
\end{figure}

\textcolor{black}{To address the problem of small samples, generative adversarial nets (GANs) used generator and discriminator in a game-like manner to obtain good performance on image applications} \cite{goodfellow2014generative,wang2021generative}. Specifically, the generator can generate new samples, according existing samples \cite{gui2021review}. The discriminator is used to  distinguish the samples from generator \cite{gui2021review}. Due to their strong learning abilities, GANs become popular image super-resolution methods \cite{bell2019blind}. \textcolor{black}{For instance, Park et al. \cite{park2023kernel} combined kernel idea and GAN to extract more structural information to enhance image sharpness and edge thickness for image super-resolution.} However, there are few studies summarizing these GANs for SISR. Also, differing from previous work based deep learning techniques for image super-resolution, i.e., Refs.\cite{jiang2021deep, anwar2020deep}, we can not only refer to importance of GANs for low- and high-levels in terms of big and small samples, but also first deeply product a summary of GANs in image super-resolution, according to combination of different training ways (i.e., supervised, semi-supervised and unsupervised manners), network architectures, prior knowledge, loss functions and multiple tasks, which can makes readers easier know principle, improvements, superiority and inferiority of different GANs for image super-resolution. That is, in this paper, we conduct a comprehensive  overview of \textcolor{black}{209} papers to show their performance, pros, cons, complexity, challenges and potential research points, etc. First, we show the effects of GANs for image applications. Second, we present popular architectures for GANs in big and small samples for image applications. Third, we analyze motivations, implementations and differences of GANs based optimization methods and discriminative learning for image super-resolution in terms of supervised, semi-supervised and unsupervised manners, where these GANs are worked by combining different network architectures, prior knowledge, loss functions and multiple tasks for image super-resolution. Fifth, we compare these GANs using experimental setting, quantitative analysis (i.e., PSNR, SSIM, complexity and running time) and qualitative analysis. Finally, we report on potential research points and existing challenges of GANs for image super-resolution. The overall architecture of this paper is shown in Fig \ref{fig:1}. 
%%%

The remainder of this survey is organized as follows: Section 2 reviews the developments of GANs; Section 3 surveys popular GANs for image applications; Section 4 focuses on introduction of existing GANs via three ways on SISR; Section 5 compares performance of mentioned GANs from Section 2 for SISR; Section 6 offers potential directions and challenges of GANs in image super-resolution; and Section 7 concludes the overview.

\section{Developments of GANs}
Traditional machine learning methods prefer to use prior knowledge to improve performance of image processing applications \cite{sun2010gradient}. For instance, Sun et al. \cite{sun2010gradient} proposed a gradient profile to restore more detailed information for improving performance of image super-resolution. Although machine learning methods based prior knowledge has fast execution speed, they have some drawbacks. First, they required manual setting parameters to achieve better performance on image tasks. Second, they required complex optimization methods to find optimized parameters. According to mentioned challenges, deep learning methods are developed \cite{bird2022fruit}. Deep learning methods used deep networks, i.e., CNNs to automatically learn features rather than manual setting parameters to obtain effective effects in image processing tasks, i.e., image classification \cite{bird2022fruit}, image inpainting \cite{liu2021pd} and image super-resolution \cite{wang2020deep}. Although these methods are effective big samples, they are limited for image tasks with small samples \cite{goodfellow2014generative}.  

\begin{figure}[!htbp]
\centering
{\includegraphics[width=3.5in]{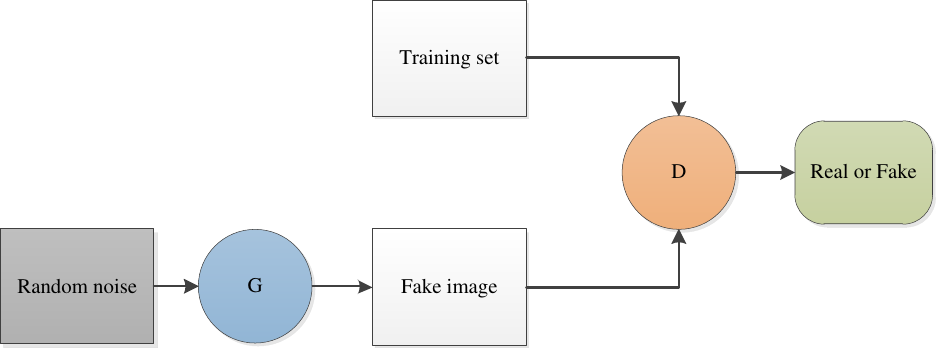}
\label{fig_second_case}}
\caption{Architecture of generative adversarial network (GAN).}
\label{fig:2}
\end{figure}

To address problems above, GANs are presented in image processing \cite{goodfellow2014generative}. GANs consist of generator network and discriminator network. The generator network is used to generate new samples, according to given samples. The discriminator network is used to determine truth of obtained new samples. When generator and discriminator is balance, a GAN model is finished. The work process of GAN can be shown in Fig \ref{fig:2}, where G and D denote a generator network and discriminator network. To better understand GANs, we introduce several basic GANs as follows.
\begin{figure}[!htbp]
\centering
{\includegraphics[width=3.5in]{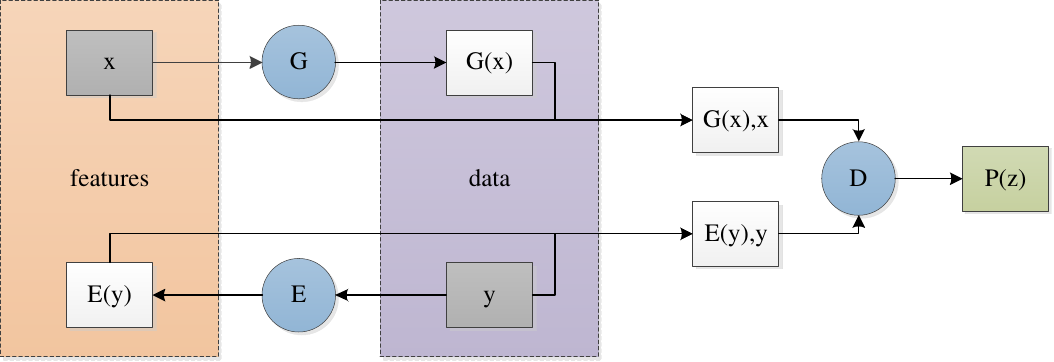}
\label{fig_second_case}}
\caption{Architecture of bidirectional generative adversarial network (BiGAN).}
\label{fig:3}
\end{figure}

To obtain more realistic effects, conditional information is fused into a GAN (CGAN) to randomly generate images, which are closer to real images \cite{mirza2014conditional}. CGAN improves GAN to obtain more robust data, which has an important reference value to GANs for computer vision applications. Subsequently, increasing the depth of GAN instead of the original multilayer perceptron in a CNN to improve expressive ability of GAN is developed for complex vision tasks \cite{radford2015unsupervised}. To mine more useful information, the bidirectional generative adversarial network (BiGAN) used dual encoders to collaborate a generator and discriminator to obtain richer information for improving performance in anomaly detection, which is shown in Fig \ref{fig:3}\cite{donahue2016adversarial}. In Fig \ref{fig:3}, x denotes a feature vector, E is an encoder and y expresses an image from discriminator.

\begin{figure}[!htbp]
\centering

{\includegraphics[width=3.5in]{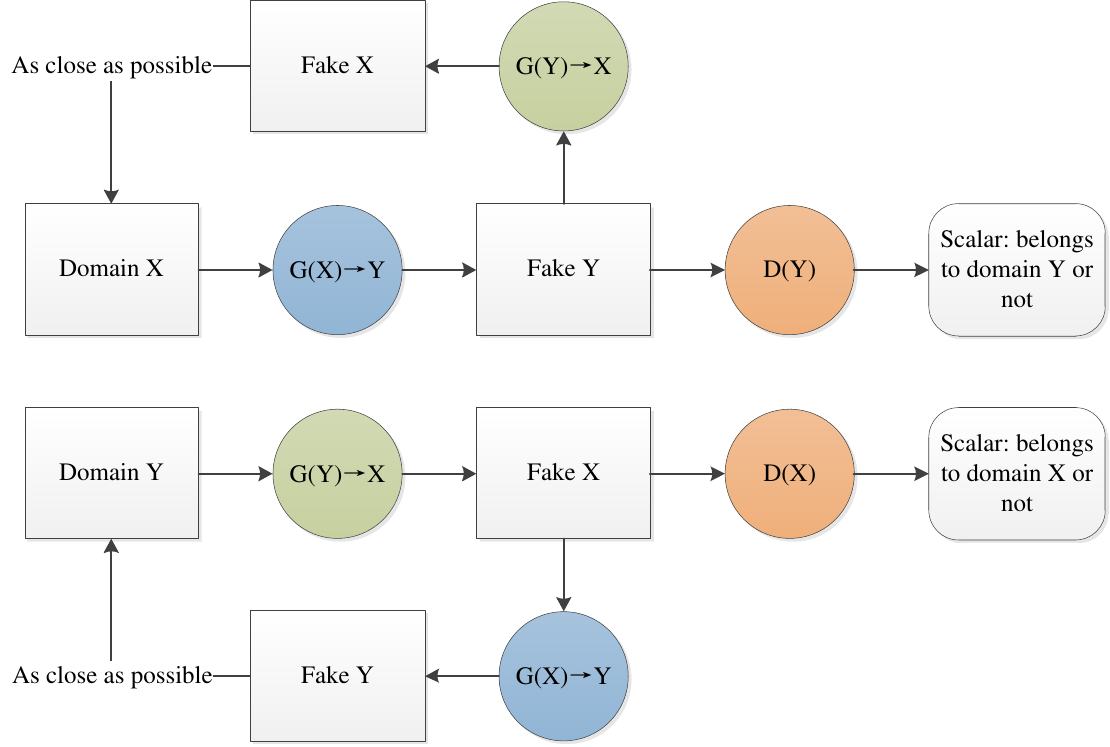}
\label{fig_second_case}}
\caption{Architecture of cycle-consistent adversarial network (CycleGAN).}
\label{fig:4}
\end{figure}

It is known that pretrained operations can be used to accelerate the training speed of CNNs for image recognition \cite{hinterstoisser2018pre}. This idea can be treated as an energy drive. Inspired by that, Zhao et al. proposed an energy-based generative adversarial network (EBGAN) by using a pretraining operation into a discriminator to improve the performance in image recognition \cite{zhao2016energy}. To keep consistency of obtained features with original images, cycle-consistent adversarial network (CycleGAN) relies on a cyclic architecture to achieve an excellent style transfer effect \cite{zhu2017unpaired} as illustrated in Fig \ref{fig:4}.

\begin{figure}[!htbp]
\centering
{\includegraphics[width=3.0in]{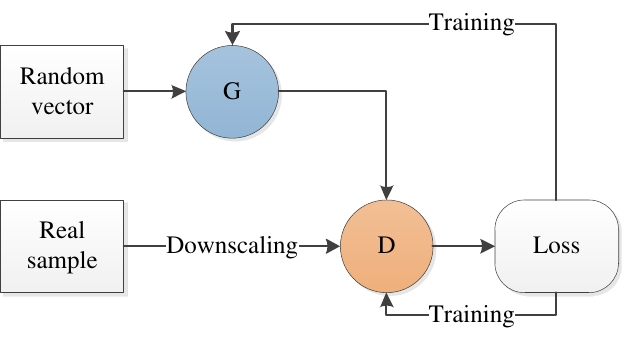}
\label{fig_second_case}}
\caption{Architecture of StyleGAN.}
\label{fig:5}
\end{figure}

\begin{figure}[!htbp]
\centering
{\includegraphics[width=3.5in]{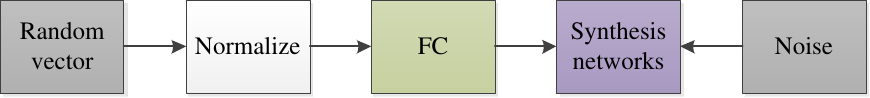}
\label{fig_second_case}}
\caption{The structure of generator in the StyleGAN.}
\label{fig:6}
\end{figure}

Although pretrained operations are useful for training efficiency of network models, they may suffer from mode collapse. To address this problem, Wasserstein GAN (WGAN) used weight clipping to enhance importance of Lipschitz constraint to improve the stability of training a GAN \cite{arjovsky2017wasserstein}. WGAN used weight clipping to perform well. However, it is easier to cause gradient vanishing or gradient exploding \cite{gulrajani2017improved}. To resolve this issue, WGAN used a gradient penalty (treated as WGAN-GP) to break the limitation of Lipschitz for pursuing good performance in computer vision applications \cite{brock2018large}. To further improve results of image generation, GAN enlarged batch size and used truncation trick as well as BIGGAN can make a tradeoff between variety and fidelity \cite{brock2018large}. To better obtained features of different parts of an image (i.e., freckles and hair), style-based GAN (StyleGAN) uses feature decoupling to control different features and finish style transfer for image generation \cite{karras2019style}. The architecture of StyleGAN and its generator are shown in Fig \ref{fig:5} and Fig \ref{fig:6}. 

In recent years, GANs with good performance have been applied in the fields of image processing, natural language processing (NLP) and video processing. Also, there are other variants based on GANs for multimedia applications, such as Laplacian pyramid of GAN (LAPGAN) \cite{denton2015deep}, coupled GAN (CoupleGAN) \cite{liu2016coupled}, self-attention GAN (SAGAN) \cite{zhang2019self}, loss-sensitive GAN (LSGAN) \cite{qi2020loss}. These methods emphasize how to generate high-quality images through various sampling mechanisms. However, researchers focused applications of GANs from 2019, i.e., FUNIT \cite{liu2019few}, SPADE \cite{park2019semantic} and U-GAT-IT \cite{kim2019u}. Illustrations of more GANs are shown in Table \ref{tab:1}.
\begin{table}
  \caption{Introduction of many GANs.}
  \centering
\scalebox{0.5}[0.6]{
\begin{tabular}{cccc}
\toprule
Models & Methods & Applications & Key words \\
\midrule
GAN \cite{goodfellow2014generative} & GAN & Image generation & GAN in a semi-supervised way for image generation \\
DICCGAN \cite{bird2022fruit} & CGAN & Image classification & Conditional GAN for image classification \\
PD-GAN \cite{liu2021pd} & GAN & Image inpainting & GAN for image inpainting and image restoration \\
CGAN \cite{mirza2014conditional} & GAN & Image generation & GAN in a supervised way for image generation \\
DCGAN \cite{radford2015unsupervised} & GAN & Image generation & GAN in an unsupervised way for image generation \\
BiGAN \cite{donahue2016adversarial} & GAN & Image generation & GAN with encoder in an unsupervised way for image generation \\
EBGAN \cite{zhao2016energy} & GAN & Image generation and training nets & GAN based energy for image generation \\
CycleGAN \cite{zhu2017unpaired} & GAN & Image generation & GAN with cycle-consistent for image generation \\
WGAN-GP \cite{gulrajani2017improved} & GAN & Image generation & GAN with gradient penalty for image generation \\
BIGGAN \cite{brock2018large} & GAN & Image super-resolution & GAN with big channels of image super-resolution \\
StyleGAN \cite{karras2019style} & GAN & Image generation & GAN with stochastic variation for image generation \\
LAPGAN \cite{denton2015deep} & CGAN & Image super-resolution & GAN with Laplacian pyramid for image super-resolution \\
CoupleGAN \cite{liu2016coupled} & GAN & Image generation & GAN for both up-sampling and image generation \\
SAGAN \cite{zhang2019self} & GAN & Image generation & Unsupervised GAN with self-attention for image generation \\
FUNIT \cite{liu2019few} & GAN & Image translation & GAN in an unsupervised way for image-to-image translation \\
SPADE \cite{park2019semantic} & GAN & Image generation & GAN with spatially-adaptive normalization for image generation \\
U-GAT-IT \cite{kim2019u} & GAN & Image translation & GAN with attention in an unsupervised way for image-to-image translation \\
\bottomrule
\end{tabular}}
\label{tab:1}
\end{table}

\section{Popular GANs for image applications}
\begin{figure}[!htbp]
\centering
{\includegraphics[width=5.5in]{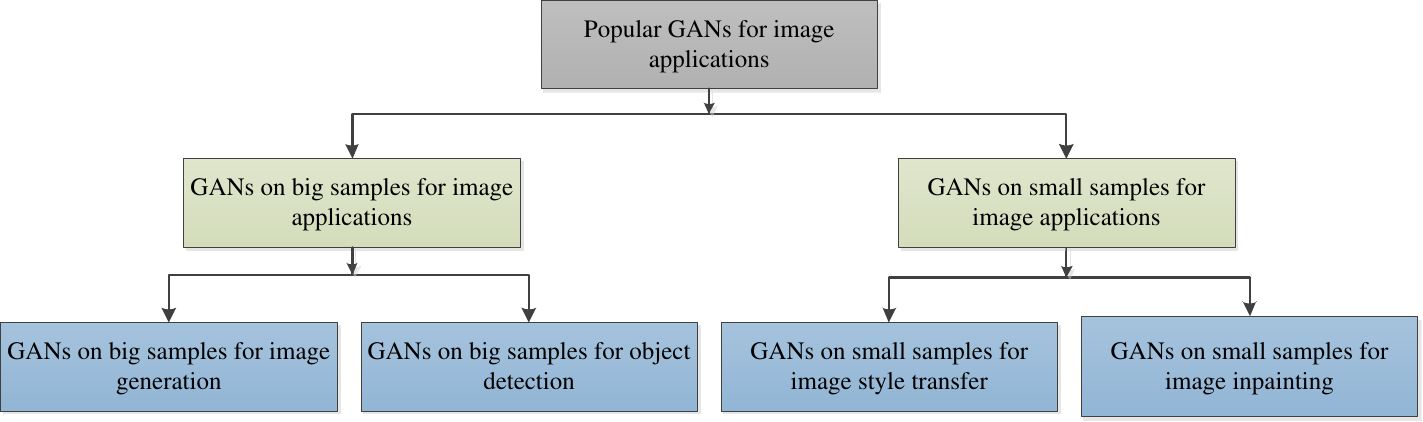}
\label{fig_second_case}}
\caption{Frame of popular GANs for image applications.}
\label{fig:7}
\end{figure}

According to mentioned illustrations, it is known that variants of GANs based on properties of vision tasks are developed in Section \uppercase\expandafter{\romannumeral2}. To further know GANs, we show different GANs on training data, i.e., big samples and small samples for different high- and low-level computer vision tasks as shown in Fig \ref{fig:7}.

\subsection{GANs on big samples for image applications}
\subsubsection{GANs on big samples for image generation}
Good performance of image generation depends on rich samples. Inspired by that, GANs are improved for image generation \cite{gui2021review}. That is, GANs use generator to produce more samples from high-dimensional data to cooperate discriminator for promoting results of image generation. For instance, boundary equilibrium generative adversarial networks (BEGAN) used obtained loss from Wasserstein to match loss of auto-encoder in the discriminator and achieve a balance between a generator and discriminator, which can obtain more texture information than that of common GANs in image generation \cite{berthelot2017began}. To control different parts of a face, StyleGAN decoupled different features to form a feature space for finishing transfer of texture information \cite{karras2019style}. \textcolor{black}{Enhanced GAN for image generation (EIGGAN) \cite{bergmann2017learning} improves the performance of the generator by incorporating a spatial attention mechanism and parallel residual operations, thereby achieving higher quality and more realistic image generation effects in large-scale image generation tasks.} Besides, texture synthesis is another important application of image generation \cite{li2016precomputed}. For instance, Markovian GANs (MGAN) can quickly capture texture date of Markovian patches to achieve function of real-time texture synthesis \cite{li2016precomputed}, where Markovian patches can be obtained Ref. \cite{gui2021review}. Periodic spatial GAN (PSGAN) \cite{bergmann2017learning} is a variant of spatial GAN (SGAN) \cite{jetchev2016texture}, which can learn periodic textures of big datasets and a single image. These methods can be summarized in Table \ref{tab:2}.

\begin{table}[htbp!]
\caption{GANs on big samples for image generation.}
\label{tab:2}
\centering
\scalebox{0.7}[0.75]{
\begin{tabular}{cccc}
\hline
Models & Methods & Key words \\
\hline \\
StyleGAN \cite{karras2019style} & GAN & GAN with stochastic variation for image generation & \\
BEGAN \cite{berthelot2017began} & GAN & GAN with upsampling for image generation & \\
\textcolor{black}{EIGGAN} \cite{tian2024enhanced} & GAN & GAN with spatial attention for image generation & \\
MGAN \cite{li2016precomputed} & GAN & GAN with Markovian for texture synthesis & \\
PSGAN \cite{bergmann2017learning} & GAN & Periodic GAN for texture synthesis & \\
SGAN \cite{jetchev2016texture} & GAN & GAN with spatial tensor for texture synthesis & \\
\hline
\end{tabular}}
\end{table}

\subsubsection{GANs on big samples for object detection}
Object detection has wide applications in the industry, i.e., smart transportation \cite{wang2021real} and medical diagnosis \cite{gerdprasert2021object}, etc. However, complex environments have huge challenges for pursuing good performance of object detection methods \cite{zou2019object}. Rich data is important for object detection. Existing methods used a data-driven strategy to collect a large-scale dataset including different object examples under different conditions to obtain an object detector. However, the obtained dataset does not contain all kinds of deformed and occluded objects, which limits effects of object detection methods. To resolve the issue, GANs are used for object detection \cite{ehsani2018segan, li2017perceptual}. Ehsani et al. used segmentation and generation in a GANs from invisible parts in the objects to overcome occluded objects \cite{ehsani2018segan}. To address a challenge of small object detection on low-resolution and noisy representation, a perceptual GAN (Perceptual GAN) reduced differences of small objects and big objects to improve performance in small object detection \cite{li2017perceptual}. That is, its generator converted poor perceived representation from small objects to high-resolution big objects to fool a discriminator, where mentioned big objects are similar to real big objects \cite{li2017perceptual}. To obtain sufficient information of objects, an end-to-end multi-task generative adversarial network (SOD-MTGAN) used a generator to recover detailed information for generating high-quality images for achieving accurate detection \cite{bai2018sod}. Also, a discriminator transferred classification and regression losses in a back-propagated way into a generator \cite{bai2018sod}. Two operations can extract objects from backgrounds to achieve good performance in object detection. More detailed information is shown in Table \ref{tab:3}. 
\begin{table}[htbp!]
\caption{GANs on big samples for object detection.}
\centering
\scalebox{0.7}[0.75]{
\begin{tabular}{cccc}
\hline
Models &Methods &Key words\\
\hline\\
SeGAN \cite{ehsani2018segan}           & GAN     & GAN with segmentor   for object detection                     &\\
Perceptual GAN   \cite{li2017perceptual} & GAN     & GAN with super-resolved representation for object detection &\\
SOD-MTGAN \cite{bai2018sod}      & GAN     & Multi-task GAN for   object detection                         &\\
\hline
\end{tabular}}
\label{tab:3}
\end{table}

\subsection{GANs on small samples for image applications}
\subsubsection{GANs on small samples for image style transfer}
Makeup has important applications in the real world \cite{yuan2022ramt}. To save costs, visual makeup software is developed, leading to image style transfer (i.e., image-to-image) translation becoming a research hotspot in the field of computer vision in recent years \cite{gui2021review}. GANs are good tools for style transfer on small samples, which can be used to establish mappings between given images and object images \cite{gui2021review}. The obtained mappings are strongly related to aligned image pairs \cite{isola2017image}. However, we found that the above mappings do not match  our ideal models in terms of transfer effects \cite{zhu2017unpaired}. Motivated by that, CycleGAN used two pairs of a generator and discriminator in a cycle consistent way to learn two mappings for achieving style transfer \cite{zhu2017unpaired}. CycleGAN had two phases in style transfer. In the first phase, an adversarial loss \cite{park2019semantic} was used to ensure the quality of generated images. In the second phase, a cycle consistency loss  \cite{zhu2017unpaired} was utilized to guarantee that predicted images to fell into the desired domains \cite{chang2018generating}. CycleGAN had the following merits. It does not require paired training examples \cite{chang2018generating}. And it does not require that the input image and the output image have the same low-dimensional embedding space \cite{zhu2017unpaired}. Due to its excellent properties, many variants of CycleGAN have been conducted for many vision tasks, i.e., image style transfer \cite{zhu2017unpaired, chen2021arcyclegan}, object transfiguration \cite{kim2018u} and image enhancement \cite{you2019fundus}, etc. More GANs on small samples for image style transfer can be found in Table~\ref{tab:4}.
\begin{table}[htbp!]
\caption{GANs on small samples for image style transfer.}
\centering
\scalebox{0.55}[0.60]{
\begin{tabular}{cccc}
\hline
Models &Methods &Key words\\
\hline\\
RAMT-GAN \cite{yuan2022ramt}  & GAN      & GAN for image   style transfer on makeup &\\
CycleGAN \cite{zhu2017unpaired}  & GAN      & Cycle-consistent   GAN for image-to-image translation &\\
CATVGAN \cite{Xie2020Correlation} & GAN      & Correlation   alignment GAN for image style transfer &\\
ITCGAN \cite{isola2017image}     & CGAN     & CGAN with U-net   for image-to-image translation &\\
ArCycleGAN \cite{chen2021arcyclegan} & GAN      & GAN with attribute   registration for image-to-image translation &\\
URCycleGAN \cite{kim2018u} & CycleGAN & CycleGAN with U-net   for image-to-image translation &\\
ECycleGAN \cite{you2019fundus} & CycleGAN & CycleGAN with convolutional   block attention module (CBAM) for image-to-image translation &\\
\hline
\end{tabular}}
\label{tab:4}
\end{table}

\subsubsection{GANs on small samples for image inpainting}
Images have played important roles in human–computer interaction in the real world \cite{cowie2001emotion}. However, they may be damaged when they were collected by digital cameras, which has a negative impact on high-level computer vision tasks. Thus, image inpainting had important values in the real world \cite{guillemot2013image}. Due to missing pixels, image inpainting suffered from enormous challenges \cite{elharrouss2020image}. To overcome shortcoming above, GANs are used to generate useful information to repair damaged images based on the surrounding pixels in the damaged images \cite{demir2018patch}. For instance, GAN used a reconstruction loss, two adversarial losses and a semantic parsing loss to guarantee pixel faithfulness and local-global contents consistency for face image inpainting \cite{li2017generative}. Although this method can generate useful information, which may cause boundary artifacts, distorted structures and blurry textures inconsistent with surrounding areas \cite{zhang2022gan, yu2018generative}. To resolve this issue, Zhang et al. embedded prior knowledge into a GAN to generate more detailed information for achieving good performance in image inpainting \cite{zhang2022gan}. Yu et al. exploited a contextual attention mechanism to improve a GAN for obtaining excellent visual effect in image inpainting \cite{yu2018generative}. Typical GANs on small samples for image inpainting is summarized in Table \ref{tab:5}.
\begin{table}[htbp!]
\caption{GANs on small samples for image inpainting.}
\centering
\scalebox{0.75}[0.8]{
\begin{tabular}{cccc}
\hline
Models &Methods &Key words\\
\hline\\
PGGAN \cite{demir2018patch} & GAN      & GAN based patch for image inpainting &\\
DE-GAN \cite{zhang2022gan} & GAN      & GAN with prior knowledge for face inpainting &\\
GFC \cite{li2017generative} & GAN      & GAN with autoencoder for image inpainting &\\
GIICA \cite{yu2018generative} & WGAN     & WGAN with attention model for image inpainting &\\
\hline
\end{tabular}}
\label{tab:5}
\end{table}

\section{GANs for image super-resolutions}
According to mentioned illustrations, it is clear that GANs have many important applications in image processing. Also, image super-resolution is crucial for high-level vision tasks, i.e., medical image diagnosis and weather forecast, etc. Thus, GANs in image super-resolution have important significance in the real world. However, there are few summaries about GANs for image super-resolution. Inspired by that, we show GANs for image super-resolution, according to supervised GANs, semi-supervised GANs and unsupervised GANs for image super-resolution as shown in Fig \ref{fig:2}. Specifically, supervised GANs in image super-resolution include supervised GANs based designed network architectures, supervised GANs based prior knowledge, supervised GANs with improved loss functions and supervised GANs based multi-tasks for image super-resolution. Semi-supervised GANs for image super-resolution contain semi-supervised GANs based designed network architectures, semi-supervised GANs with improved loss functions and semi-supervised GANs based multi-tasks for image super-resolution. 

Unsupervised GANs for image super-resolution consists of unsupervised GANs based designed network architectures, unsupervised GANs based prior knowledge, unsupervised GANs with improved loss functions and unsupervised GANs based multi-tasks in image super-resolution. More information of GANs on image super-resolution can be illustrated as follows.

\subsection{Supervised GANs for image super-resolution}
\subsubsection{\textcolor{black}{Supervised GANs based designed network architectures for image super-resolution}}

GANs in a supervised way to train image super-resolution models are very mainstream. Also, designing GANs via improving network architectures are very novel. Thus, improved GANs in a supervised way for image super-resolution are very popular. That can improve GANs by designing novel discriminator networks, generator networks, attributes of image super-resolution task, complexity and computational costs. For example, Laplacian pyramid of adversarial networks (LAPGAN) fused a cascade of convolutional networks into Laplacian pyramid network in a coarse-to-fine way \textcolor{black}{to obtain high-quality images for assisting image recognition task} \cite{denton2015deep}. \textcolor{black}{To overcome the effect of big scales,} curvature and highlight compact regions can be used to obtain a local salient map for adapting big scales in image-resolution \cite{mahapatra2017image}. More research on improving discriminators and generators is shown as follows.  

\textcolor{black}{In terms of designing novel and discriminators and generators}, progressive growing generative adversarial networks (PGGAN or ProGAN) utilized different convolutional layers to progressively enlarge low-resolution images to improve image qualities for image recognition \cite{karras2017progressive}. \textcolor{black}{To achieve better visual quality with more realistic and natural textures,} %% zxy add
an enhanced SRGAN (ESRGAN) used residual dense blocks into a generator without batch normalization to mine more detailed information for image super-resolution \cite{wang2018esrgan}. \textcolor{black}{To eliminate effects of checkerboard artifacts and the unpleasing high-frequency}, multi-discriminators were proposed for image super-resolution \cite{lee2019multi}. \textcolor{black}{That is, a perspective discriminator was used to overcome checkerboard artifacts and a gradient perspective was utilized to address unpleasing high-frequency question in image super-resolution.} \textcolor{black}{To improve the perceptual quality of predicted images}, ESRGAN+ fused two adjacent layers in a residual learning way based on residual dense blocks in a generator to enhance memory abilities and added noise in a generator to obtain stochastic variation and obtain more details of high-resolution images \cite{rakotonirina2020esrgan+}. 

\begin{figure}[htbp]
  \centering
  \includegraphics[width=0.55\linewidth]{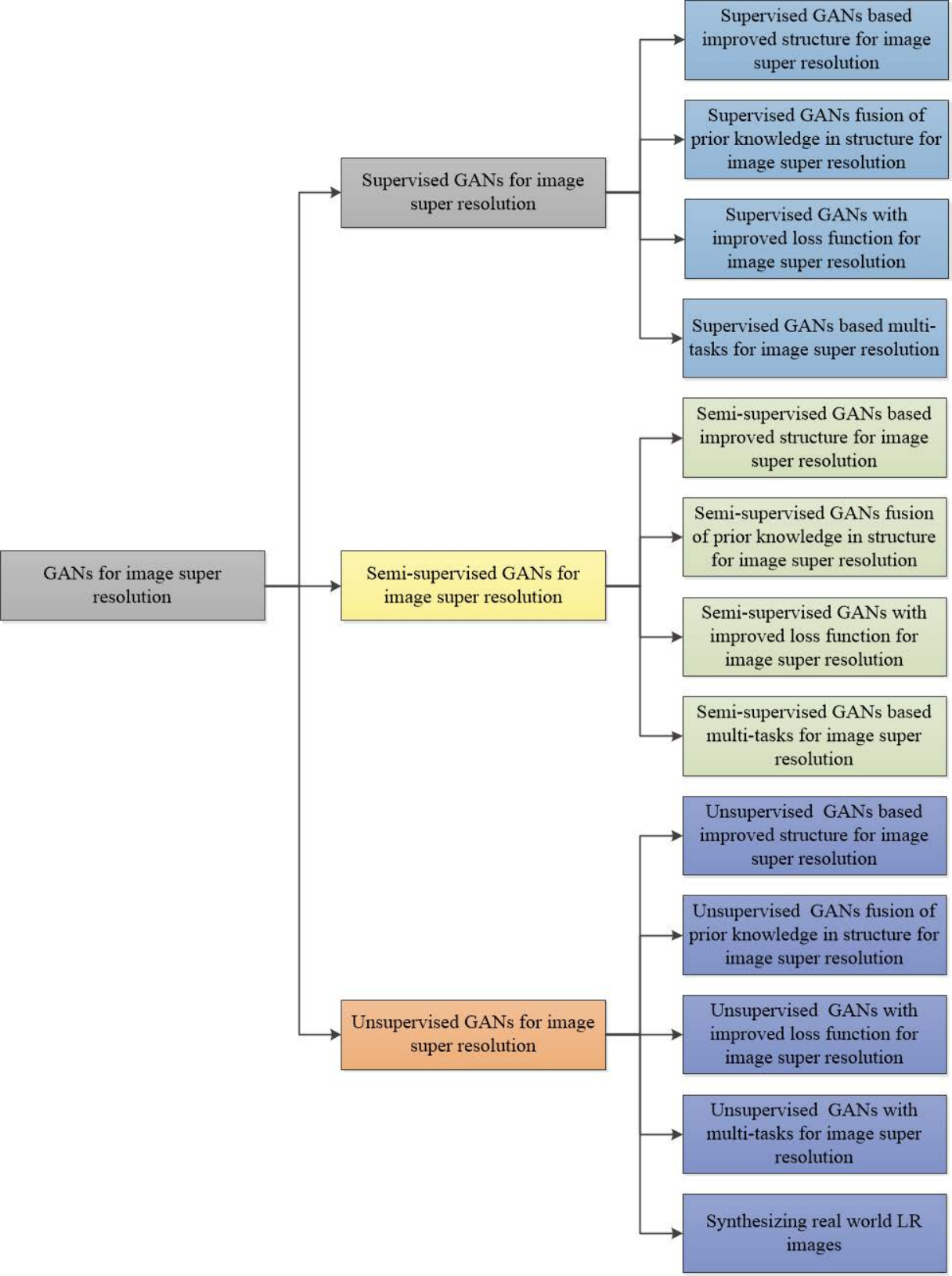}
  \caption{Frame of GANs for image super-resolution.}
  \label{fig:8}
\end{figure}
   
Restoring detailed information may generate artifacts, which can seriously affect qualities of restored images \cite{zhang2020supervised}. \textcolor{black}{The methods mentioned in this paragraph can effectively alleviate this phenomenon. }%% zxy add
\textcolor{black}{In terms of face image super-resolution,} Zhang et al. used a supervised pixel-wise GAN (SPGAN) to obtain higher-quality face images via given low-resolution face images of multiple scale factors to remove artifacts in image super-resolution \cite{zhang2020supervised}. \textcolor{black}{In terms of remote sensing image super-resolution,} Gong et al. used enlighten blocks to make a deep network achieve a reliable point and used self-supervised hierarchical perceptual loss to overcome effects of artifacts in remote sensing image super-resolution \cite{gong2021enlighten}. Dharejo et al. used Wavelet Transform (WT) characteristics into a transferred GAN \textcolor{black}{to eliminate artifacts to improve quality of predicted remote sensing images} \cite{dharejo2021twist}. Moustafa et al. embedded squeeze-and-excitation blocks and residual blocks into a generator \textcolor{black}{to obtain more high-frequency details} \cite{moustafa2021satellite}. \textcolor{black}{Besides, Wasserstein distance is used to enhance the stability of training a remote sensing super-resolution model} \cite{moustafa2021satellite}. \textcolor{black}{To address pseudo-textures problem,} a saliency analysis is fused with a GAN to obtain a salient map that can be used to distinguish difference between a discriminator and a generator \cite{ma2019sd}. 

\textcolor{black}{To obtain more detailed information in image super-resolution}, a lot of GANs are developed \cite{ko2021multi}. Ko et al. used Laplacian idea and edge in a GAN \textcolor{black}{to obtain more useful information to improve clarities of predicted face images} \cite{ko2021multi}. Using tensor structures in a GAN can facilitate texture information for SR \cite{ding2019tgan}. Using multiple generators in a GAN \textcolor{black}{can obtain more realistic texture details, which was useful to recover high-quality images} \cite{zareapoor2019diverse, xing2021deep}. \textcolor{black}{To obtain better visual effects}, a gradually GAN used gradual growing factors in a GAN to improve performance in SISR \cite{shamsolmoali2019g}.

\textcolor{black}{To reduce computational costs and memory}, Ma et al. used two-stage generator in a supervision way \textcolor{black}{to extract more effective features of cytopathological images, which can reduce the cost of data acquisition and save cost} \cite{ma2020pathsrgan}. Cheng et al. designed a generator by multi-scale feature aggregation and a discriminator via a PatchGAN to \textcolor{black}{reduce memory consumption} for a GAN on SR \cite{cheng2021mfagan}. Besides, distilling a generator and discriminator can \textcolor{black}{accelerate the training efficiency} of a GAN model for SR \cite{cheng2021mfagan}. More supervised GANs for image super-resolution are shown in Table \ref{tab:6}.

\begin{table}
  \caption{Supervised GANs for image super-resolution in section 2.1.1.}
  \centering
\scalebox{0.6}[0.75]{
\begin{tabular}{cccc}
\toprule
Models &Methods &Key words\\
\midrule
LAPGAN \cite{denton2015deep} & CGAN           & CGAN with Laplacian   Pyramid for image super-resolution                                                         &\\
LSMGAN \cite{mahapatra2017image} & CGAN           & CGAN with local   saliency maps for retinal image super-resolution                                               &\\
PGGAN \cite{karras2017progressive} & GAN            & Progressive   growing GAN for image super-resolution                                                             &\\
ESRGAN \cite{wang2018esrgan} & SRGAN & SRGAN with Residual-in-Residual   Dense Block (RRDB) and relativistic discriminator for image super-resolution   &\\
MPDGAN \cite{lee2019multi} & GAN            & GAN with multi-discriminators   for image super-resolution                                                       &\\
ESRGAN+ \cite{rakotonirina2020esrgan+} & ESRGAN         & ESRGAN with Residual-in-Residual   Dense Residual Block (RRDRB) for image super-resolution                       &\\
SPGAN \cite{zhang2020supervised} & GAN            & GAN with identity-based   discriminator for face image super-resolution                                          &\\
MLGE \cite{ko2021multi} & LAPGAN         & LAPGAN with edge   information for face image super-resolution                                                   &\\
SD-GAN \cite{ma2019sd} & GAN            & GAN for remote   sensing image super-resolution                                                                  &\\
PathSRGAN \cite{ma2020pathsrgan} & SRGAN          & SRGAN with RRDB   for cytopathology image super-resolution                                                       &\\
Enlighten-GAN   \cite{gong2021enlighten} & GAN            & GAN with enlighten   block for remote sensing image super-resolution                                             &\\
TWIST-GAN \cite{dharejo2021twist} & GAN            & GAN with wavelet   transform (WT) for remote sensing image super-resolution                                      &\\
SCSE-GAN \cite{moustafa2021satellite} & GAN            & GAN with SCSE   block for image super-resolution                                                                 &\\
MFAGAN \cite{cheng2021mfagan} & GAN            & GAN with multi-scale   feature aggregation net for image super-resolution                                        &\\
TGAN \cite{ding2019tgan} & GAN            & GAN with visual   tracking and attention networks for image super-resolution                                     &\\
DGAN \cite{zareapoor2019diverse} & GAN            & GAN with disentangled   representation learning and anisotropic BRDF reconstruction for image   super-resolution &\\
DMGAN \cite{xing2021deep} & GAN            & GAN with two same   generators for image super-resolution                                                        &\\
G-GANISR \cite{shamsolmoali2019g} & GAN            & GAN with gradual learning for image   super-resolution                                                           &\\
SRGAN \cite{ledig2017photo} & GAN            & GAN with deep ResNet   for image super-resolution                                                                &\\
RaGAN \cite{jolicoeur2018relativistic} & GAN            & GAN with relativistic   discriminator for image super-resolution                                                 &\\
LE-GAN \cite{2021arXiv211108685s} & GAN            & GAN with a latent   encoder for realistic hyperspectral image super-resolution                                   &\\
NCSR \cite{kim2021noise} & GAN            & GAN with a noise   conditional layer for image super-resolution                                                  &\\
Beby-GAN \cite{li2021best} & GAN            & GAN with a   region-aware adversarial learning strategy for image super-resolution                               &\\
MA-GAN \cite{xu2021multi} & GAN            & GAN with pyramidal   convolution for image super-resolution                                                      &\\
CMRI-CGAN \cite{xia2021super} & CGAN           & CGAN with optical   flow component for magnetic resonance image super-resolution &                      \\
D-SRGAN \cite{demiray2021d} & SRGAN          & SRGAN for image   super-resolution &                      \\
LMISR-GAN \cite{ma2021medical} & GAN            & GAN with residual   channel attention block for medical image super-resolution                                   &\\
\bottomrule
\end{tabular}}
\label{tab:6}
\end{table}

\subsubsection{Supervised GANs based prior knowledge for image super-resolution}
\textcolor{black}{It is known that combination of discriminative method and optimization can make a tradeoff between efficiency and performance} \cite{zhang2017learning}. \textcolor{black}{To handle the blind distortions in real low resolution images,} Guan et al. used high-resolution image to low-resolution image network and low-resolution image to high-resolution image network with nearest neighbor down-sampling method to learn detailed information and noise prior for image super-resolution \cite{guan2019srdgan}. Chan et al. used \textcolor{black}{rich and diverse priors in a given pretrained to mine latent representative information for generating realistic textures for image super-resolution} \cite{chan2021glean}. Liu et al. used a gradient prior into a GAN \textcolor{black}{to suppress the effect of blur kernel estimation for image super-resolution} \cite{liu2019infrared}.
  
\subsubsection{Supervised GANs with improved loss functions for image super-resolution}
Loss function can affect performance and efficiency of a trained SR model. Thus, we analyze the combination of GANs with different loss functions in image super-resolution \cite{zhang2019ranksrgan}. Zhang et al. trained a Ranker \textcolor{black}{to obtain representation of perceptual metrics} and used a rank-content loss in a GAN to \textcolor{black}{improve visual effects in image super-resolution} \cite{zhang2019ranksrgan}. \textcolor{black}{To eliminate effect of artifacts}, Zhu et al. used image quality assessment metric to implement a novel loss function \textcolor{black}{to enhance the stability for image super-resolution} \cite{zhu2020gan}. \textcolor{black}{To decrease complexity of GAN model in image super-resolution}, Fuoli et al. used a Fourier space supervision loss to recover lost high-frequency information to improve predicted image quality and accelerate training efficiency in SISR \cite{fuoli2021fourier}. \textcolor{black}{To enhance stability of a SR model}, using residual blocks and a self-attention layer in a GAN enhances robustness of a trained SR model. Also, combining improved Wasserstein gradient penalty and perceptual Loss \textcolor{black}{enhances stability of a SR model} \cite{shahidi2021breast}. \textcolor{black}{To extract accurate features}, fusing a measurement loss function into a GAN can obtain more detailed information to obtain clearer images \cite{shahsavari2021proposing}. 

\subsubsection{Supervised GANs based multi-tasks for image super-resolution}
Improving image quality is important for high-level vision tasks, i.e., image recognition \cite{talab2019super}. Besides, devices often suffer from effects of multiple factors, i.e., device hardware, camera shakes and shooting distances, which results in collected images are damaged. That may include noise and low-resolution pixels. Thus, \textcolor{black}{addressing the multi-tasks for GANs are very necessary} \cite{hu2019rtsrgan}. For instance, Adil et al. exploited SRGAN and a denoising module to obtain a clear image. Then, they used a network \textcolor{black}{to learn unique representative information for identifying a person} \cite{adil2020multi}. In terms of image super-resolution and object detection, Wang et al. used multi-class cyclic super-resolution GAN to restore high-quality images, and \textcolor{black}{used a YOLOv5 detector to finish object detection task} \cite{wang2022remote}. Zhang et al. used a fully connected network to implement a generator for obtaining high-definition plate images and a multi-task discriminator is used \textcolor{black}{to enhance super-resolution and recognition tasks} \cite{zhang2018joint}. The use of an adversarial learning was a good tool \textcolor{black}{to simultaneously address text recognition and super-resolution} \cite{xu2020srr}.  

In terms of complex damaged image restoration, GANs are good choices \cite{li2022multi}. For instance, Li et al. used a multi-scale residual block and an attention mechanism in a GAN to remove noise and restore detailed information in CTA image super-resolution \cite{li2022multi}. Nneji et al. improved a VGG19 to fine-tune two sub-networks with a wavelet technique to simultaneously address COVID-19 image denoising and super-resolution problems \cite{nneji2022fine}. More information is shown in Table \ref{tab:7}.
\begin{table}
  \caption{Supervised GANs for image super-resolution in section 2.1.2 to section 2.1.4.}
  \centering
\scalebox{0.7}[0.8]{
\begin{tabular}{cccc}
\toprule
Models &Methods &Key words\\
\midrule
SRDGAN \cite{guan2019srdgan} & GAN    & GAN with GMSR for   image super-resolution                                                      &\\
GLEAN \cite{chan2021glean} & GAN    & GAN with   pre-trained models for image super-resolution                                        &\\
I-SRGAN \cite{liu2019infrared} & GAN    & GAN with infrared prior   knowledge for image super-resolution on infrared image                &\\
RankSRGAN \cite{zhang2019ranksrgan} & SRGAN  & SRGAN with ranker   for image super-resolution                                                  &\\
GMGAN \cite{zhu2020gan} & GAN    & GAN with a novel   quality loss for image super-resolution                                      &\\
FSLSR \cite{fuoli2021fourier} & GAN    & GAN with fourier   space losses for image super-resolution                                      &\\
I-WAGAN \cite{shahidi2021breast} & GAN    & GAN with improved   wasserstein gradient penalty and perceptual loss for image super-resolution &\\
CESR-GAN \cite{shahsavari2021proposing} & GAN    & GAN with a   feature-based measurement loss function for image super-resolution                 &\\
RTSRGAN \cite{hu2019rtsrgan} & SRGAN  & SRGAN for real   time image super-resolution                                                    &\\
MSSRGAN \cite{adil2020multi} & ESRGAN & ESRGAN with denoising module for image super-resolution                                         &\\
RSISRGAN \cite{wang2022remote} & GAN    & GAN for image   super-resolution on RSI                                                         &\\
JPLSRGAN \cite{zhang2018joint} & GAN    & GAN for license plate recognition and   image super-resolution                                  &\\
SRR-GAN \cite{xu2020srr} & GAN    & GAN for image super-resolution   on text images                                                 &\\
MRD-GAN \cite{li2022multi} & GAN    & GAN with attention   mechanism for image super-resolution and denoising                        & \\
MESRGAN+ \cite{nneji2022fine} & ESRGAN & ESRGAN with siamese   network for image super-resolution and denoising.                        &\\
RealESRGAN \cite{wang2021real} & ESRGAN & ESRGAN with pure synthetic   data for blind image super-resolution &\\
SNPE-SRGAN \cite{tampubolon2021snpe} & SRGAN  & SRGAN with SPNE   for image super-resolution                                                    &\\
SOUP-GAN \cite{zhang2022soup} & GAN    & GAN with 3D MRI   for image super-resolution &\\
\bottomrule
\end{tabular}}
\label{tab:7}
\end{table}

\subsection{Semi-supervised GANs for image super-resolution}
\subsubsection{\textcolor{black}{Semi-supervised GANs based designed network architectures for image super-resolution}}
\textcolor{black}{For real problems with less data, semi-supervised techniques are developed}. For instance, asking patients takes multiple CT scans with additional radiation doses to conduct paired CT images for training SR models in clinical practice is not realistic. Motivated by that, GANs in semi-supervised ways are used for image super-resolution \cite{you2019ct}. \textcolor{black}{For instance, by maintaining the cycle-consistency of Wasserstein distance,} a mapping from noisy low-resolution images to high-resolution images was built \cite{you2019ct}. Besides, combining a convolutional neural network, residual learning operations in a GAN can facilitate more detailed information for image super-resolution \cite{you2019ct}. \textcolor{black}{To resolve super-resolution with few labeled samples}, Xia et al. used soft multi-labels to implement a semi-supervised super-resolution method for person re-identification \cite{xia2021real}. That is, first, a GAN is used to conduct a SR model. Second, a graph convolutional network is exploited to construct relationship of local features from a person. Third, some labeled samples are used to train unlabeled samples via a graph convolutional network.  

\subsubsection{Semi-supervised GANs with improved loss functions and semi-supervised GANs based multi-tasks for image super-resolution}
The combinations of semi-supervised GANs and loss functions are also effective in image super-resolution \cite{jiang2020novel}. For example, Jiang et al. combined an adversarial loss, a cycle-consistency loss, an identity loss and a joint sparsifying transform loss into a GAN in a semi-supervised way to train a CT image super-resolution model \cite{jiang2020novel}. \textcolor{black}{Although this model made a significantly progress on some evaluation criteria, it was still disturbed by artifacts and noise.}

In terms of multi-tasks, Nicolo et al. proposed to use a mixed adversarial Gaussian domain adaptation in a GAN in a semi-supervised way to obtain more useful information for implementing a 3D super-resolution and segmentation \cite{savioli2021joint}. \textcolor{black}{This method achieved better performance on many metrics.} More information of semi-supervised GANs in image super-resolution can be illustrated in Table \ref{tab:8}. 
\begin{table}
  \caption{Semi-supervised GANs for image super-resolution in section 2.2.}
  \centering
\scalebox{0.6}[0.7]{
\begin{tabular}{cccc}
\toprule
Models &Methods &Key words\\
\midrule
GAN-CIRCLE \cite{you2019ct} & GAN & GAN with cycle-consistency   of Wasserstein distance in a semi-supervised way for noisy image   super-resolution            &\\
MSSR \cite{xia2021real} & GAN & GAN with soft   multi-labels in a semi-supervised way for image super-resolution                                            &\\
CTGAN \cite{jiang2020novel} & GAN & GAN with four   losses in a semi-supervised way for image super-resolution                                                  &\\
Gemini-GAN \cite{savioli2021joint} & GAN & GAN with mixed   adversarial Gaussian domain adaptation in a semi-supervised way for 3D   super-resolution and segmentation &\\
\bottomrule
\end{tabular}}
  \label{tab:8}
\end{table}

\subsection{Unsupervised GANs for image super-resolution}
\textcolor{black}{Collected images in the real world have less pairs, \textcolor{black}{which suffers from challenge for supervised GANs in SISR}. To address this phenomenon, unsupervised GANs are presented} \cite{yuan2018unsupervised}. Unsupervised GANs for image super-resolution can be divided into \textcolor{black}{six} types: \textcolor{black}{ improved architectures, prior knowledge, loss functions, multi-task learning, real image super-resolution and real-world LR image synthesis in GAN-based unsupervised super-resolution, as follows.}

\subsubsection{\textcolor{black}{Unsupervised GANs based designed network architectures for image super-resolution}}
CycleGANs have obtained success in unsupervised ways in image-to-image translation applications \cite{zhu2017unpaired}. \textcolor{black}{Accordingly, the CycleGANs are extended into SISR to address unpair images (i.e., low-resolution and high-resolution) in the datasets in real world} \cite{yuan2018unsupervised}. Yuan et al. used a CycleGAN for blind super-resolution over the following phases \cite{yuan2018unsupervised}. The first phase removed noise from noisy and low-resolution images. The second phase resorted to an up-sampled operation in a pre-trained deep network to enhance the obtained low-resolution images. The third phase used a fine-tune mechanism for a GAN to obtain high-resolution images. \textcolor{black}{To address blind super-resolution,} bidirectional structural consistency was used into a GAN in an unsupervised way to train a blind SR model and construct high-quality images \cite{zhao2018unsupervised}. Alternatively, Zhang et al. exploited multiple GANs as basis components to implement an improved CycleGAN for train an unsupervised SR model \cite{zhang2019multiple}. \textcolor{black}{And achieves comparable performance with many state-of-the-art supervised models.}

There are also other popular methods that use GANs in unsupervised ways for image super-resolution \cite{prajapati2020unsupervised}. \textcolor{black}{To improve the learning ability of a SR model in the real world}, it combines an unsupervised learning and a mean opinion score in a GAN to improve perceptual quality in the real world image super-resolution \cite{prajapati2020unsupervised}. \textcolor{black}{To break fixed downscaling kernel}, \textcolor{black}{KernelGAN \cite{bell2019blind}, TVG-KernelGAN\cite{park2023kernel}, and Internal-GAN \cite{shocher2018ingan} are used} to obtain an internal distribution of patches in the blind image super-resolution. \textcolor{black}{To accelerate the training speed}, a guidance module was used in a GAN to quickly seek a correct mapping from a low-resolution domain to a high-resolution domain in unpaired image super-resolution \cite{lian2019fg}. \textcolor{black}{To improve the accuracy of medical diagnosis}, Song et al. used dual GANs in a self-supervised way to mine high dimensional information for PET image super-resolution \cite{mahapatra2017image}. Besides, other SR methods can have an important reference value for unsupervised GANs with for image super-resolution. For example, \textcolor{black}{to improve both image synthesis quality and representation learning performance under the unsupervised setting,} %% zxy add
Wang et al. used an unsupervised method to translate real low-resolution images to real low-resolution images \cite{wang2020transformation}. \textcolor{black}{To achieve better degradation learning and super-resolution performance,} Chen et al. resorted to a supervised super-resolution method to convert obtained real low-resolution images into real high-resolution images \cite{chen2020unsupervised}. More information of mentioned unsupervised GANs for image super-resolution can be shown in Table \ref{tab:9} as follows. 
\begin{table}
  \caption{Unsupervised GANs based designed network architectures for image super-resolution.}
  \centering
\scalebox{0.7}[0.75]{
\begin{tabular}{cccc}
\toprule
Models &Methods &Key words\\
\midrule
CinCGAN \cite{yuan2018unsupervised} & GAN          & Unsupervised GAN   for image super-resolution                                                 &\\
DNSR \cite{zhao2018unsupervised} & GAN          & Unsupervised GAN   with bidirectional structural consistency for blind image super-resolution &\\
MCinCGAN \cite{zhang2019multiple} & CycleGAN     & Unsupervised GAN   for image super-resolution                                                 &\\
\textcolor{black}{RWSR-CycleGAN}   \cite{kim2020unsupervised} & CycleGAN     & Unsupervised GAN   for image super-resolution                                                 &\\
USISResNet \cite{prajapati2020unsupervised} & GAN          & Unsupervised GAN   with USISResNet for image super-resolution                                 &\\
\textcolor{black}{ULRWSR} \cite{lugmayr2019unsupervised} & GAN          & Unsupervised GAN   with pixel wise supervision for image super-resolution                     &\\
KernelGAN \cite{bell2019blind} & Internal-GAN & Unsupervised GAN   for blind image super-resolution                                           &\\
InGAN \cite{shocher2018ingan} & GAN          & Unsupervised GAN   for image super-resolution                                                 &\\
FG-SRGAN \cite{lian2019fg} & SRGAN        & Unsupervised GAN   with a guided block for image super-resolution                             &\\
PETSRGAN \cite{mahapatra2017image} & GAN          & Unsupervised GAN   with a self-supervised   way for PET image super-resolution                &\\
TrGAN \cite{wang2020transformation} & GAN          & Unsupervised GAN   for image synthesis and super-resolution                                   &\\
CycleSR \cite{chen2020unsupervised} & GAN          & Unsupervised GAN with   an indirect supervised path for image super-resolution                &\\
UGAN-Circle \cite{guha2021unsupervised} & GAN-Circle   & Unsupervised   GAN-Circle for image super-resolution on CT images &\\
\bottomrule
\end{tabular}}
  \label{tab:9}
\end{table}

\subsubsection{Unsupervised GANs based prior knowledge for image super-resolution}

Combining unsupervised GANs and prior knowledge in unsupervised GANs can better address unpair image super-resolution \cite{lin2018deep}.  Lin et al. combined data error, a regular term and an adversarial loss \textcolor{black}{to guarantee consistency of local-global content and pixel faithfulness in a GAN} in an unsupervised way to train an image super-resolution model \cite{lin2018deep}. \textcolor{black}{To better support medical diagnosis}, Das et al. combined adversarial learning in a GAN, cycle consistency and prior knowledge, i.e., identity mapping prior to facilitate more useful information i.e., spatial correlation, color and texture information \textcolor{black}{for obtaining cleaner high-quality images} \cite{das2020unsupervised}. \textcolor{black}{In terms of remoting sensing super-resolution,} a random noise is used in a GAN to reconstruct satellite images \cite{wang2021enhanced}. Then, authors conducted image prior by transforming the reference image into a latent space \cite{wang2021enhanced}. Finally, they updated the noise and latent space to transfer obtained structure information and texture information \textcolor{black}{for improving resolution of remote sensing images} \cite{wang2021enhanced}.

\subsubsection{Unsupervised GANs with improved loss functions for image super-resolution}

Combining loss functions and GANs in an unsupervised way is useful for training image super-resolution models in the real world \cite{zhang2020unsupervised}. For instance, Zhang et al. used a novel loss function based image quality assessment in a GAN to obtain accurate texture information and more visual effects \cite{zhang2020unsupervised}. Besides, an encoder-decoder architecture is embedded in this GAN \textcolor{black}{to mine more structure information for pursuing high-quality images} of a generator from this GAN \cite{zhang2020unsupervised}. Han et al. depended on SAGAN and L1 loss in a GAN in an unsupervised manner \textcolor{black}{to act multi-sequence structural MRI for detecting braining anomalies} \cite{han2021madgan}. Also, Zhang et al. fused a content loss into a GAN in an unsupervised manner \textcolor{black}{to improve SR results of hyperspectral images} \cite{zhang2021degradation}. Unsupervised GANs based prior knowledge and improved loss functions for image super-resolution can be summarized in Table \ref{tab:10}. 
\begin{table}
  \caption{Unsupervised GANs based prior knowledge and improved loss functions for image super-resolution.}
  \centering
\scalebox{0.75}[0.8]{
\begin{tabular}{cccc}
\toprule
Models &Methods &Key words\\
\midrule
DULGAN \cite{lin2018deep} & GAN   & GAN with prior for   image super-resolution                                               &\\
USROCTGAN \cite{das2020unsupervised} & GAN   & GAN with cycle   consistency and identity mapping priors for image super-resolution       &\\
EIPGAN \cite{wang2021enhanced} & GAN   & GAN with remote   sensing image prior for image super-resolution                          &\\
URSGAN \cite{zhang2020unsupervised} & GAN   & GAN with prior   based image quality assessment for remote sensing image super-resolution &\\
MADGAN \cite{han2021madgan} & SAGAN & SAGAN with L1 loss   for medical image super-resolution                                   &\\
DLGAN \cite{zhang2021degradation} & GAN   & GAN with a content   loss for hyperspectral image super-resolution &\\
\bottomrule
\end{tabular}}
\label{tab:10}
\end{table}

\begin{table}
  \caption{Unsupervised GANs based multi-tasks for image super-resolution.}
  \label{tab:11}
  \centering
\scalebox{0.7}[0.8]{
\begin{tabular}{cccc}
\toprule
Models &Methods &Key words\\
\midrule
VAEGAN \cite{prajapati2021unsupervised} & GAN & GAN with variational   auto-encoder and quality assessment idea for image super-resolution and   denoising &\\
ASLGAN \cite{cui2022unsupervised} & GAN & GAN with   low-pass-filter loss and weighted MR images for MRI image super-resolution   and denoising      &\\
Pix2NeRF \cite{cai2022pix2nerf} & GAN & Optimizing a   periodic implicit GAN for 3D-aware image synthesis and super-resolution                     &\\
Pi-GAN \cite{chan2021pi} & GAN & GAN with periodic   activation functions for 3D-aware image synthesis and image super-resolution &\\
\bottomrule
\end{tabular}}
\end{table}

\subsubsection{Unsupervised GANs based multi-tasks for image super-resolution}
Unsupervised GANs are good tools to address multi-tasks, i.e., noisy low-resolution image super-resolution. For instance, Prajapati et al. transferred a variational auto-encoder and the idea of quality assessment in a GAN \textcolor{black}{to deal with image denoising and SR tasks} \cite{prajapati2021unsupervised}. Cui et al. relied on low-pass-filter loss and weighted MR images in a GAN in an unsupervised GAN \textcolor{black}{to mine texture information for removing noise and recovering resolution of MRI images} \cite{cui2022unsupervised}. Cai et al. presented a pipeline that optimizes a periodic implicit GAN \textcolor{black}{to obtain neural radiance fields for image synthesis and image super-resolution based on 3D} \cite{cai2022pix2nerf}. More unsupervised GANs based multi-tasks for image super-resolution can be presented in Table \ref{tab:11}.

\subsubsection{\textcolor{black}{Unsupervised GANs for real image super-resolution}}
%无监督GAN位于图像超分辨%
\textcolor{black}{An important application of GAN-based SR models is dealing with real world LR images. To eliminate checkerboard artifacts, an upsampling module containing a bilinear interpolation and a transposed convolution was used in an unsupervised CycleGAN to improve visual effects of restored images in the real world \cite{kim2020unsupervised}. To recover more natural image characteristics, Lugmayr et al. combined unsupervised and supervised ways for blind image super-resolution \cite{lugmayr2019unsupervised}. The first step learned to invert the effects of bicubic down sampling operation in a GAN in an unsupervised way to extract useful information from natural images \cite{lugmayr2019unsupervised}. To generate image pairs in the real world, the second step used a pixel-wise network in a supervised way to obtain high-resolution images \cite{lugmayr2019unsupervised}. 
To remove unpleasing noise and artifacts, Park et al. \cite{park2023content} divided the large and complicated distribution of real-world images into smaller subsets based on similar contents. Then, they learned the various contents and separable features via multiple discriminators \cite{park2023content}. In terms of computed tomography (CT) images, Li et al. \cite{li20234} utilized noise injection and a GAN to generate realistic low-resolution images to build training image pairs. This data preprocessing operation enables the proposed KerGAN to generate more precise details and better perceptual quality medical images \cite{li20234}.} %% zxy add

%%2.3.6这一段可以删掉，对应审稿意见3.7-1
\subsubsection{\textcolor{black}{Synthesizing real world LR images}}
\textcolor{black}{An important application of GAN-based SR models is dealing with real world LR images. So, how to obtain real image pairs with different scale factors to be used as training data becomes a problem that we must solve. Starting from SRCNN \cite{dong2015image}, there are three main ways regarding the preparation of super-resolution datasets: synthetic data super resolution, blind super-resolution and real image super-resolution. Among them, synthetic data super-resolution is the most common way of data preparation, and this route usually uses the intuitive interpolation method as a degradation kernel for downsampling to obtain the corresponding training image pairs. However, in the face of complex degradation factors in the real world, synthetic data often do not generalize well. Therefore, on the basis of the interpolation method, blind super-resolution models add more real world degradation factors, i.e., noise, blur, compression etc. to the degradation kernel to improve the performance of real world applications \cite{wang2021real, ren2020real}. Finally, the best theoretical performance is achieved by directly utilizing a camera to capture real world images of the same scene at different magnifications. However, in this process, the problem of lens distortion is encountered which makes the image acquisition difficult. Currently, the publicly available real image datasets are RealSR \cite{cai2019toward} and DRealSR \cite{wei2020component}.}
\begin{table}[htbp]
  \caption{\textcolor{black}{Datasets (i.e., training datasets and test datasets) of GANs for image super-resolution.}}
  \label{tab:12}
  \centering
\scalebox{0.55}[0.55]{
\begin{tabular}{cccll}
\toprule
Training datasets &Test datasets\\
\midrule
\begin{tabular}[c]{@{}l@{}}
CIFAR10 \cite{krizhevsky2009learning}, \\   STL \cite{wang2016unsupervised}, \\    LSUN \cite{zhang2015large}, \\  ImageNet \cite{russakovsky2015imagenet}, \\  Celeb A \cite{liu2015deep},\\   DIV2K \cite{agustsson2017ntire},\\  Flickr2K \cite{wang2019flickr1024},\\  OST \cite{wang2018recovering},\\  CAT \cite{zhang2008cat},\\  Market-1501 \cite{zheng2015scalable},   \\    Duke MTMC-reID   \cite{ristani2016performance, zheng2017unlabeled},\\  GeoEye-1 satellite   dataset \cite{ma2019sd},\\  Whole slide images   (WSIs) \cite{ma2020pathsrgan},\\  MNIST \cite{lecun1998gradient}, \\  PASCAL2  \cite{everingham2010pascal},\\  Set5 \cite{bevilacqua2012low}, \\  Set14 \cite{zeyde2010single}, \\  BSD100 \cite{martin2001database}, \\  Urban100 \cite{huang2015single},\\Tibia Dataset   \cite{chen2018quantitative}, \\ Abdominal Dataset   \cite{mccollough2020low},\\ CUHK03 \cite{li2014deepreid}, \\ MSMT17 \cite{wei2018person},\\ LUNA \cite{setio2017validation}, \\ Data Science Bowl   2017 (DSB) \cite{kuan2017deep},\\ UKDHP \cite{woodbridge2013mridb}, \\ SG \cite{woodbridge2013mridb}, \\ UKBB \cite{bycroft2018uk},\\Widerface \cite{yang2016wider},\\ NTIRE-2020 Real-world   SR Challenge \cite{lugmayr2020ntire}, \\ KADID-10K \cite{lin2019kadid},\\ DPED \cite{ignatov2017dslr}, \\ DF2K \cite{wang2018esrgan}, \\ NTIRE’2018   Blind-SR challenge \cite{timofte2018ntire}, \\ DIV2K random   kernel (DIV2KRK) \cite{agustsson2017ntire},\\ LS3D-W \cite{bulat2017far},\\ CELEBA-HQ \cite{karras2017progressive}, \\ LSUN-BEDROOM \cite{yu2015lsun},   \\ NTIRE 2020 \cite{lugmayr2020ntire},\\ 91-images \cite{yang2010image}, \\ Berkeley   segmentation \cite{martin2001database}, \\ BSDS500 \cite{yang2016object},\\ Training datasets   of USROCTGAN \cite{fang2013fast, fang2012sparsity}, \\ SD-OCT dataset   \cite{fang2012sparsity},\\ UC Merced dataset   \cite{yang2010bag}, \\ NWPU-RESIS45   \cite{cheng2017remote}, \\ WHU-RS19 \cite{dai2010satellite}.
\end{tabular} 
& \begin{tabular}[c]{@{}l@{}}
CIFAR10 \cite{krizhevsky2009learning}, \\   STL \cite{wang2016unsupervised}, \\  LSUN \cite{zhang2015large},\\  Set5 \cite{bevilacqua2012low}, \\  Set14 \cite{zeyde2010single}, \\  BSD100 \cite{martin2001database},\\  CELEBA \cite{liu2015deep}, \\  CAT \cite{zhang2008cat},\\  OST300 \cite{wang2018recovering}, \\  the PIRM datasets   \cite{blau20182018},\\ Market-1501 \cite{zheng2015scalable},\\  GeoEye-1 satellite   dataset \cite{ma2019sd},\\  WSIs \cite{ma2020pathsrgan},\\  MNIST \cite{lecun1998gradient}, \\  PASCAL2 \cite{everingham2010pascal},\\Tibia Dataset   \cite{chen2018quantitative}, \\ Abdominal Dataset   \cite{mccollough2020low},\\  CUHK03 \cite{li2014deepreid}\\ LUNA \cite{setio2017validation}, \\ DSB \cite{kuan2017deep},\\ SG \cite{woodbridge2013mridb}, \\ UKBB \cite{bycroft2018uk},\\Urban100 \cite{huang2015single}, \\ NTIRE 2020 Real World SR challenge \cite{lugmayr2020ntire},\\ DPED \cite{ignatov2017dslr}, \\ DIV2KRK \cite{agustsson2017ntire},\\ Widerface \cite{yang2016wider},\\ ImageNet \cite{russakovsky2015imagenet}, \\ CELEBA-HQ \cite{karras2017progressive}, \\ LSUN-BEDROOM \cite{yu2015lsun},\\ Tests datasets of   USROCTGAN \cite{fang2013fast},\\ Test datasets of USRGAN   \cite{zhang2020unsupervised}.
\end{tabular} &\\ 
\bottomrule
\end{tabular}}
\end{table}

\section{Comparing performance of GANs for image super-resolution}
To make readers conveniently know GANs in image super-resolution, we compare
super-resolution performance of these GANs from datasets and experimental settings to quantitative and qualitative analysis in this section. More information can be shown as follows.

\subsection{Datasets}
Mentioned GANs can be divided into three kinds: supervised methods, semi-supervised methods and unsupervised methods for image super-resolution, which make datasets have three categories, training datasets and test datasets for supervised methods, semi-supervised methods and unsupervised methods. These datasets can be summarized as follows. 

(1)	Supervised GANs for image-resolution

Training datasets: CIFAR10 \cite{krizhevsky2009learning}, STL \cite{wang2016unsupervised}, LSUN \cite{zhang2015large}, ImageNet \cite{russakovsky2015imagenet}, Celeb A \cite{liu2015deep}, DIV2K \cite{agustsson2017ntire}, Flickr2K \cite{wang2019flickr1024}, OST \cite{wang2018recovering}, CAT \cite{zhang2008cat} Market-1501 \cite{zheng2015scalable}, Duke MTMC-reID \cite{ristani2016performance, zheng2017unlabeled}, GeoEye-1 satellite dataset \cite{ma2019sd}, Whole slide images (WSIs) \cite{ma2020pathsrgan}, MNIST \cite{lecun1998gradient} and PASCAL2 \cite{everingham2010pascal}.

Test datasets: CIFAR10 \cite{krizhevsky2009learning}, STL \cite{wang2016unsupervised}, LSUN \cite{zhang2015large}, Set5 \cite{bevilacqua2012low}, Set14 \cite{zeyde2010single}, BSD100 \cite{martin2001database}, CELEBA \cite{liu2015deep}, OST300 \cite{zhang2015large}, CAT \cite{zhang2008cat}, PIRM datasets \cite{blau20182018}, Market-1501 \cite{zheng2015scalable}, GeoEye-1 satellite dataset \cite{ma2019sd}, WSIs \cite{ma2020pathsrgan} MNIST \cite{lecun1998gradient} and PASCAL2 \cite{everingham2010pascal}.

(2)	Semi-supervised GANs for image-resolution

Training datasets: Market-1501 \cite{zheng2015scalable}, Tibia Dataset \cite{chen2018quantitative}, Abdominal Dataset \cite{mccollough2020low}, CUHK03 \cite{li2014deepreid}, MSMT17 \cite{wei2018person}, LUNA \cite{setio2017validation}, Data Science Bowl 2017 (DSB) \cite{kuan2017deep}, UKDHP \cite{woodbridge2013mridb}, SG \cite{woodbridge2013mridb} and UKBB \cite{bycroft2018uk}.

Test datasets: Tibia Dataset \cite{chen2018quantitative}, Abdominal Dataset \cite{mccollough2020low}, CUHK03 \cite{li2014deepreid}, Widerface \cite{yang2016wider}, LUNA \cite{setio2017validation}, DSB \cite{kuan2017deep}, SG \cite{woodbridge2013mridb} and UKBB \cite{bycroft2018uk}.

(3)	Unsupervised GANs for image-resolution

Training datasets: CIFAR10 \cite{krizhevsky2009learning}, ImageNet \cite{russakovsky2015imagenet}, DIV2K \cite{agustsson2017ntire}, DIV2K random kernel (DIV2KRK) \cite{agustsson2017ntire}, Flickr2K \cite{wang2019flickr1024}, Widerface \cite{yang2016wider}, NTIRE 2020 Real World SR challenge \cite{lugmayr2020ntire}, KADID-10K \cite{lin2019kadid}, DPED \cite{ignatov2017dslr}, DF2K \cite{wang2018esrgan}, NTIRE’ 2018 Blind-SR challenge \cite{timofte2018ntire}, LS3D-W \cite{bulat2017far}, CELEBA-HQ \cite{karras2017progressive}, LSUN-BEDROOM \cite{yu2015lsun}, ILSVRC2012 \cite{russakovsky2015imagenet, nguyen2017plug}, NTIRE 2020 \cite{lugmayr2020ntire}, 91-images \cite{yang2010image}, Berkeley segmentation \cite{martin2001database}, BSDS500 \cite{yang2016object}, Training datasets of USROCTGAN \cite{fang2013fast, fang2012sparsity}, SD-OCT dataset \cite{fang2012sparsity}, UC Merced dataset \cite{yang2010bag}, NWPU-RESIS45 \cite{cheng2017remote} and WHU-RS19 \cite{dai2010satellite}.

Test datasets: CIFAR10 \cite{krizhevsky2009learning}, ImageNet \cite{russakovsky2015imagenet}, Set5 \cite{bevilacqua2012low}, Set14 \cite{zeyde2010single}, BSD100 \cite{martin2001database}, DIV2K \cite{agustsson2017ntire}, DIV2KRK \cite{agustsson2017ntire}, Urban100 \cite{huang2015single}, Widerface \cite{yang2016wider}, NTIRE 2020 \cite{lugmayr2020ntire}, NTIRE 2020 Real-world SR Challenge \cite{lugmayr2020ntire}, NTIRE 2020 Real World SR challenge validation dataset \cite{lugmayr2020ntire}, DPED \cite{ignatov2017dslr}, CELEBA-HQ \cite{karras2017progressive}, LSUN-BEDROOM \cite{yu2015lsun}, Test datasets of USROCTGAN \cite{fang2013fast, fang2012sparsity}, and Test datasets of USRGAN \cite{zhang2020unsupervised}.

These mentioned datasets about GANs for image super-resolution can be shown in Table \ref{tab:12}. To make readers easier understand datasets of different methods via different GANs for different training ways on image super-resolutions, we conduct Table \ref{tab:13} to show their detailed information.

\begin{table}
  \caption{Different GANs on image super-resolution for different training ways.}
  \label{tab:13}
  \centering
\scalebox{0.42}[0.5]{
\begin{tabular}{ccccc}
\toprule
Training ways &Methods &Training datasets &Test datasets\\
\midrule
Supervised ways
& \begin{tabular}[c]{@{}c@{}}
LAPGAN \cite{denton2015deep} \\   SRGAN \cite{ledig2017photo} \\ PGGAN \cite{karras2017progressive} \\  ESRGAN \cite{wang2018esrgan} \\  RaGAN \cite{jolicoeur2018relativistic}\\   ESRGAN+ \cite{rakotonirina2020esrgan+}\\  SPGAN \cite{zhang2020supervised}\\  SD-GAN \cite{ma2019sd} \\  PathSRGAN \cite{ma2020pathsrgan}\\  TGAN \cite{ding2019tgan}   \\    DGAN \cite{zareapoor2019diverse}\\  G-GANISR \cite{shamsolmoali2019g}
\end{tabular}

& \begin{tabular}[c]{@{}c@{}}
CIFAR10 \cite{krizhevsky2009learning}, STL   \cite{wang2016unsupervised}, LSUN \cite{zhang2015large}\\
ImageNet \cite{russakovsky2015imagenet}\\
CIFAR10 \cite{krizhevsky2009learning},   Celeb A \cite{liu2015deep}\\
DIV2K \cite{agustsson2017ntire},   Flickr2K \cite{wang2019flickr1024}, OST \cite{wang2018recovering}\\
CIFAR10 \cite{krizhevsky2009learning}, CAT   \cite{zhang2008cat}\\
DIV2K \cite{agustsson2017ntire}\\
Market-1501 \cite{zheng2015scalable},   Duke MTMC-reID \cite{ristani2016performance, zheng2017unlabeled}\\
GeoEye-1 satellite   dataset \cite{ma2019sd}\\
Whole slide images   (WSIs) \cite{ma2020pathsrgan}\\
MNIST \cite{lecun1998gradient},   PASCAL2 \cite{everingham2010pascal}, CIFAR10 \cite{krizhevsky2009learning}\\
DIV2K \cite{agustsson2017ntire}\\
Set5 \cite{bevilacqua2012low}, Set14   \cite{zeyde2010single}, BSD100 \cite{martin2001database}, Urban100 \cite{huang2015single}\\

\end{tabular}

& \begin{tabular}[c]{@{}c@{}}
CIFAR10 \cite{krizhevsky2009learning}, STL\cite{wang2016unsupervised}, LSUN \cite{zhang2015large}\\
Set5 \cite{bevilacqua2012low}, Set14\cite{bevilacqua2012low}, BSD100 \cite{martin2001database}\\
CELEBA \cite{liu2015deep}, LSUN   \cite{zhang2015large}, CIFAR10 \cite{krizhevsky2009learning} \\
Set5 \cite{bevilacqua2012low}, Set14   \cite{zeyde2010single}, BSD100 \cite{martin2001database}, Urban100 \cite{huang2015single} \\
CAT \cite{zhang2008cat} \\
BSD100 \cite{martin2001database},   Urban100 \cite{huang2015single}, OST300 \cite{wang2018recovering}, Set5 \cite{bevilacqua2012low}, Set14   \cite{bevilacqua2012low}, the PIRM datasets \cite{blau20182018} \\
Market-1501 \cite{zheng2015scalable} \\
GeoEye-1 satellite   dataset \cite{ma2019sd} \\
WSIs \cite{ma2020pathsrgan} \\
MNIST \cite{lecun1998gradient},   PASCAL2 \cite{everingham2010pascal}, CIFAR10 \cite{krizhevsky2009learning} \\
Set5 \cite{bevilacqua2012low}, Set14   \cite{zeyde2010single}, CIFAR-10 \cite{krizhevsky2009learning}, BSD100 \cite{martin2001database} \\
Set5 \cite{bevilacqua2012low}, Set14   \cite{zeyde2010single}, Urban100 \cite{huang2015single} \\ 

\end{tabular} \\

\midrule
Semi-supervised ways
& \begin{tabular}[c]{@{}c@{}}
GAN-CIRCLE \cite{you2019ct} \\   MSSR \cite{xia2021real} \\ CTGAN \cite{jiang2020novel} \\  Gemini-GAN \cite{savioli2021joint}
\end{tabular}

& \begin{tabular}[c]{@{}c@{}}
Tibia Dataset   \cite{chen2018quantitative}, Abdominal Dataset \cite{mccollough2020low}\\
Market-1501 \cite{zheng2015scalable},   CUHK03 \cite{li2014deepreid}, MSMT17 \cite{wei2018person}\\
LUNA \cite{setio2017validation}, Data   Science Bowl 2017 (DSB) \cite{kuan2017deep}\\
UKDHP \cite{woodbridge2013mridb}, SG   \cite{woodbridge2013mridb}, UKBB \cite{bycroft2018uk}\\

\end{tabular}

& \begin{tabular}[c]{@{}c@{}}
Tibia Dataset   \cite{chen2018quantitative}, Abdominal Dataset \cite{mccollough2020low} \\
Market-1501 \cite{zheng2015scalable},   CUHK03 \cite{li2014deepreid}\\
LUNA \cite{setio2017validation}, DSB \cite{kuan2017deep} \\
SG \cite{woodbridge2013mridb}, UKBB   \cite{bycroft2018uk} \\

\end{tabular} \\

\midrule
Unsupervised ways
& \begin{tabular}[c]{@{}c@{}}
CinCGAN \cite{yuan2018unsupervised} \\   DNSR \cite{zhao2018unsupervised} \\ MCinCGAN \cite{zhang2019multiple} \\  RWSR-CycleGAN   \cite{kim2020unsupervised}\\USISResNet \cite{prajapati2020unsupervised}\\ULRWSR \cite{lugmayr2019unsupervised}\\KernelGAN \cite{bell2019blind}\\FG-SRGAN \cite{lian2019fg}\\TrGAN \cite{wang2020transformation}\\PDCGAN \cite{miyato2018cgans}\\CycleSR \cite{chen2020unsupervised}\\DULGAN \cite{lin2018deep}\\USROCTGAN \cite{das2020unsupervised}\\URSGAN \cite{zhang2020unsupervised}\\
\end{tabular}

& \begin{tabular}[c]{@{}c@{}}
DIV2K \cite{agustsson2017ntire}\\
DIV2K \cite{agustsson2017ntire},   Flickr2K \cite{wang2019flickr1024}, Widerface \cite{yang2016wider}\\
DIV2K \cite{agustsson2017ntire} \\
NTIRE 2020 Real   World SR challenge \cite{lugmayr2020ntire}\\
NTIRE-2020 Real-world   SR Challenge validation dataset \cite{lugmayr2020ntire}, DIV2K \cite{agustsson2017ntire}, Flickr2k \cite{wang2019flickr1024}, KADID-10K   \cite{lin2019kadid}\\
DPED \cite{ignatov2017dslr}, DF2K   \cite{wang2018esrgan}, DIV2K \cite{agustsson2017ntire}, Flickr2K \cite{wang2019flickr1024}\\
NTIRE’2018   Blind-SR challenge \cite{timofte2018ntire}, DIV2K random kernel (DIV2KRK) \cite{agustsson2017ntire}\\
LS3D-W \cite{bulat2017far}\\
CIFAR10 \cite{krizhevsky2009learning}, ImageNet   \cite{russakovsky2015imagenet}, CELEBA-HQ \cite{karras2017progressive}, LSUN-BEDROOM \cite{yu2015lsun}\\
ImageNet \cite{russakovsky2015imagenet},   ILSVRC2012 \cite{russakovsky2015imagenet, nguyen2017plug}, CIFAR10 \cite{krizhevsky2009learning}, CIFAR-100 \cite{krizhevsky2009learning}\\
DIV2K \cite{agustsson2017ntire}, NTIRE   2020 \cite{lugmayr2020ntire}\\
91-images \cite{yang2010image},   Berkeley segmentation \cite{martin2001database}, BSDS500 \cite{yang2016object}\\
Training datasets of USROCTGAN \cite{fang2013fast, fang2012sparsity}, SD-OCT dataset \cite{fang2012sparsity}\\
UC Merced dataset   \cite{yang2010bag}, NWPU-RESIS45 \cite{cheng2017remote}, WHU-RS19 \cite{dai2010satellite}\\
\end{tabular}

& \begin{tabular}[c]{@{}c@{}}
DIV2K \cite{agustsson2017ntire} \\
Set5 \cite{bevilacqua2012low}, Set14   \cite{zeyde2010single}, Urban100 \cite{huang2015single}, BSD100 \cite{martin2001database}, DIV2K \cite{agustsson2017ntire}\\
DIV2K \cite{agustsson2017ntire} \\
NTIRE 2020 Real   World SR challenge \cite{lugmayr2020ntire} \\
NTIRE-2020 Real-world   SR Challenge validation dataset \cite{lugmayr2020ntire}\\
DPED \cite{ignatov2017dslr}, DIV2K   \cite{agustsson2017ntire} \\
DIV2KRK \cite{agustsson2017ntire} \\
Widerface \cite{yang2016wider} \\
CIFAR10 \cite{krizhevsky2009learning}, ImageNet   \cite{russakovsky2015imagenet}, CELEBA-HQ \cite{karras2017progressive}, LSUN-BEDROOM \cite{yu2015lsun} \\
ImageNet \cite{russakovsky2015imagenet} \\
DIV2K \cite{agustsson2017ntire}, NTIRE   2020 \cite{lugmayr2020ntire} \\
Set5 \cite{bevilacqua2012low}, Set14 \cite{zeyde2010single} \\Test datasets of   USROCTGAN \cite{fang2013fast} \\Test datasets of   USRGAN \cite{zhang2020unsupervised} \\
\end{tabular} 

\\ \bottomrule
\end{tabular}}
\end{table}

\subsection{Environment configurations}
In this section, we compare the differences of environment configurations between different GANs via different training ways (i.e., supervised, semi-supervised and unsupervised ways) for image super-resolution, which contain batch size, scaling factors, deep learning framework, learning rate and iteration. That can make readers easier to conduct experiments with GANs for image super-resolution. Their information can be listed as shown in Table \ref{tab:14} as follows.

\subsection{Experimental results}
To make readers understand the performance of different GANs on image super-resolution, we use quantitative analysis and qualitative analysis to evaluate super-resolution effects of these GANs. Quantitative analysis is PSNR and SSIM of different methods via three training ways on different datasets for image super-resolution, running time and complexities of different GANs on image super-resolution. Qualitative analysis is used to evaluate qualities of recovered images.

\subsubsection{Quantitative analysis of different GANs for image super-resolution}
We use SRGAN \cite{ledig2017photo}, PGGAN \cite{karras2017progressive}, ESRGAN \cite{wang2018esrgan}, ESRGAN+ \cite{rakotonirina2020esrgan+}, DGAN \cite{zareapoor2019diverse}, G-GANISR \cite{shamsolmoali2019g}, GMGAN \cite{zhu2020gan}, SRPGAN \cite{ledig2017photo}, DNSR \cite{zhao2018unsupervised}, DULGAN \cite{lin2018deep}, CinCGAN \cite{yuan2018unsupervised},MCinCGAN \cite{zhang2019multiple}, USISResNet\cite{prajapati2020unsupervised}, ULRWSR \cite{lugmayr2019unsupervised}, KernelGAN \cite{bell2019blind}  and  CycleSR \cite{chen2020unsupervised} in one training way from supervised, semi-supervised and unsupervised ways on a public dataset from Set14 \cite{zeyde2010single},  BSD100  \cite{martin2001database} and DIV2K \cite{agustsson2017ntire} to test performance for different scales in image super-resolution as shown in Table \ref{tab:15}. For instance, ESRGAN \cite{wang2018esrgan} outperforms SRGAN \cite{ledig2017photo} in terms of PSNR and SSIM in a supervised ways on Set 14 for ×2, which shows that ESRGAN has obtained better super-resolution performance for ×2. More information of these GANs can be shown in Table \ref{tab:15}.

\begin{table}
  \caption{Environment configurations of different GANs for image super-resolution.}
  \centering
\scalebox{0.65}[0.65]{
\begin{tabular}{ccccccc}
\toprule
Training ways &Methods &Batchsize &Scaling factors &Framework &Learning rates &Iteration
\\\midrule
Supervised ways
& \begin{tabular}[c]{@{}c@{}}
LAPGAN \cite{denton2015deep} \\   ESRGAN \cite{wang2018esrgan} \\  ESRGAN+ \cite{rakotonirina2020esrgan+}\\  SPGAN \cite{zhang2020supervised}\\  SD-GAN \cite{ma2019sd} \\  TGAN \cite{ding2019tgan}   \\    DGAN \cite{zareapoor2019diverse}\\  G-GANISR \cite{shamsolmoali2019g}\\GMGAN \cite{zhu2020gan}
\end{tabular}

& \begin{tabular}[c]{@{}c@{}}
16\\16\\16\\16\\9\\32\\-\\-\\16
\end{tabular}

& \begin{tabular}[c]{@{}c@{}}
×2\\×4\\×4\\×2, ×4, ×8 and ×16\\×3\\×2\\×4, ×6 and ×8\\×4, ×6 and ×8\\×4
\end{tabular}

& \begin{tabular}[c]{@{}c@{}}
PyTorch \cite{paszke2019pytorch}\\PyTorch \cite{paszke2019pytorch}\\PyTorch \cite{paszke2019pytorch}\\TensorFlow \cite{goldsborough2016tour}\\TensorFlow \cite{goldsborough2016tour}\\TensorFlow \cite{goldsborough2016tour}\\TensorFlow \cite{goldsborough2016tour}\\PyTorch \cite{paszke2019pytorch} and TensorFlow \cite{goldsborough2016tour}\\PyTorch \cite{paszke2019pytorch}
\end{tabular} 

& \begin{tabular}[c]{@{}c@{}}
0.02\\2×$10^{-4}$, $10^{-4}$\\$10^{-4}$\\$10^{-3}$\\$10^{-3}$\\$10^{-4}$\\0.125\\$10^{-4}$\\2×$10^{-4}$
\end{tabular}

& \begin{tabular}[c]{@{}c@{}}
-\\200K, 300K\\300K\\-\\2K\\500\\25K\\800K\\100K
\end{tabular}

\\\midrule
Semi-supervised ways
& \begin{tabular}[c]{@{}c@{}}
MSSR \cite{xia2021real} \\ PSSR \cite{maeda2020unpaired}\\  CTGAN \cite{jiang2020novel}
\end{tabular}

& \begin{tabular}[c]{@{}c@{}}
64\\16\\16
\end{tabular}

& \begin{tabular}[c]{@{}c@{}}
×4\\×4\\×4
\end{tabular}

& \begin{tabular}[c]{@{}c@{}}
PyTorch \cite{paszke2019pytorch}\\PyTorch \cite{paszke2019pytorch}\\TensorFlow \cite{goldsborough2016tour}
\end{tabular}

& \begin{tabular}[c]{@{}c@{}}
$10^{-3}$, 0.01\\$10^{-4}$\\$10^{-4}$
\end{tabular}

& \begin{tabular}[c]{@{}c@{}}
-\\300K\\60K
\end{tabular}

\\\midrule
Unsupervised ways
& \begin{tabular}[c]{@{}c@{}}
CinCGAN \cite{yuan2018unsupervised} \\   DNSR \cite{zhao2018unsupervised} \\ MCinCGAN \cite{zhang2019multiple} \\  RWSR-CycleGAN   \cite{kim2020unsupervised}\\USISResNet \cite{prajapati2020unsupervised}\\ULRWSR \cite{lugmayr2019unsupervised}\\KernelGAN \cite{bell2019blind}\\InGAN \cite{shocher2018ingan}\\PETSRGAN \cite{mahapatra2017image}\\PDCGAN \cite{miyato2018cgans}\\USROCTGAN \cite{das2020unsupervised}\\URSGAN \cite{zhang2020unsupervised}\\
\end{tabular}

& \begin{tabular}[c]{@{}c@{}}
16\\16\\1\\16\\32\\-\\-\\1\\10, 20\\-\\64\\64
\end{tabular}

& \begin{tabular}[c]{@{}c@{}}
×4\\×2, ×4\\×2, ×4 and ×8\\×4\\×4\\×4\\×2, ×4\\×2\\×4\\×4\\×4\\×2, ×4
\end{tabular}

& \begin{tabular}[c]{@{}c@{}}
PyTorch \cite{paszke2019pytorch}\\TensorFlow \cite{goldsborough2016tour}\\-\\TensorFlow \cite{goldsborough2016tour}\\PyTorch \cite{paszke2019pytorch}\\TensorFlow \cite{goldsborough2016tour}\\PyTorch \cite{paszke2019pytorch}\\-\\PyTorch \cite{paszke2019pytorch}\\TensorFlow \cite{goldsborough2016tour}\\TensorFlow \cite{goldsborough2016tour}\\PyTorch \cite{paszke2019pytorch}\\
\end{tabular}

& \begin{tabular}[c]{@{}c@{}}
$10^{-4}$\\$10^{-4}$\\2×$10^{-4}$\\$10^{-4}$\\$10^{-4}$\\$10^{-4}$, 2×$10^{-4}$\\2×$10^{-4}$\\$10^{-4}$\\2×$10^{-4}$\\2×$10^{-4}$\\2×$10^{-4}$\\5×$10^{-4}$
\end{tabular}

& \begin{tabular}[c]{@{}c@{}}
400K\\106\\500K\\300K\\120K\\50K\\3K\\20K\\-\\850K\\68K\\-
\end{tabular}

\\\bottomrule
\end{tabular}}
\label{tab:14}
\end{table}

\begin{table}[htbp!]
\caption{PSNR and SSIM of different GANs via different training ways on Set14, BSD100 and DIV2K for image super-resolution.}
\label{tab:15}
\centering
\scalebox{0.8}[0.8]{
\begin{tabular}{cccccc}
\toprule
Training ways &Methods &Datasets &Scale &PSNR &SSIM\\
\midrule
\multirow{32}{*}{Supervised ways}   & \multirow{2}{*}{SRGAN \cite{ledig2017photo} } & \multirow{16}{*}{Set14 \cite{zeyde2010single}}  & ×2 & 32.14  & 0.8860 \\
                                    &                                      &                                    & ×4 & 26.02  & 0.7379 \\
                                    & \multirow{2}{*}{PGGAN \cite{karras2017progressive}}      &                                    & ×6 & 29.54  & 0.8301 \\
                                    &                                      &                                    & ×8 & 28.14  & 0.8094 \\
                                    & \multirow{2}{*}{ESRGAN \cite{wang2018esrgan}}     &                                    & ×2 & 33.62  & 0.9150 \\
                                    &                                      &                                    & ×4 & 30.50  & 0.7620 \\
                                    & ESRGAN+ \cite{rakotonirina2020esrgan+}                     &                                    & ×4 & 19.79  & -      \\
                                    & \multirow{3}{*}{DGAN \cite{zareapoor2019diverse}}       &                                    & ×4 & 31.62  & 0.9166 \\
                                    &                                      &                                    & ×6 & 28.62  & 0.9003 \\
                                    &                                      &                                    & ×8 & 26.85  & 0.8911 \\
                                    & \multirow{3}{*}{G-GANISR \cite{shamsolmoali2019g}}   &                                    & ×4 & 29.67  & -      \\
                                    &                                      &                                    & ×6 & 30.56  & 0.8881 \\
                                    &                                      &                                    & ×8 & 28.07  & 0.8803 \\
                                    & GMGAN \cite{zhu2020gan}                      &                                    & ×4 & 26.37  & 0.7055 \\
                                    & \multirow{2}{*}{SRPGAN \cite{ledig2017photo}}    &                                    & ×6 & 26.19  & 0.8187 \\
                                    &                                      &                                    & ×8 & 29.17  & 0.8733 \\\cline{2-6}
                                    & \multirow{2}{*}{SRGAN \cite{ledig2017photo}}      & \multirow{13}{*}{BSD100 \cite{martin2001database}} & ×2 & 31.89  & 0.8760 \\
                                    &                                      &                                    & ×4 & 25.16  & 0.6688 \\
                                    & \multirow{2}{*}{ESRGAN \cite{wang2018esrgan}}     &                                    & ×2 & 31.99  & 0.8870 \\
                                    &                                      &                                    & ×4 & 27.69  & 0.7120 \\
                                    & \multirow{3}{*}{DGAN \cite{zareapoor2019diverse}}       &                                    & ×4 & 31.53  & 0.9105 \\
                                    &                                      &                                    & ×6 & 29.62  & 0.8937 \\
                                    &                                      &                                    & ×8 & 27.85  & 0.8811 \\
                                    & \multirow{3}{*}{G-GANISR \cite{shamsolmoali2019g}}   &                                    & ×4 & 28.12  & -      \\
                                    &                                      &                                    & ×6 & 31.23  & 0.9273 \\
                                    &                                      &                                    & ×8 & 29.18  & 0.9065 \\
                                    & GMGAN \cite{zhu2020gan}                      &                                    & ×4 & 25.46  & 0.6592 \\
                                    & \multirow{2}{*}{SRPGAN \cite{ledig2017photo}}    & & ×6 & 27.48  & 0.8652 \\
                                    &                                      &                                    & ×8 & 23.18  & 0.8504 \\\cline{2-6}
                                    & \multirow{2}{*}{SRGAN \cite{ledig2017photo}}      & \multirow{3}{*}{DIV2K \cite{agustsson2017ntire}}   & ×2 & 25.08  & 0.7007 \\
                                    &                                      &                                    & ×4 & 28.09  & 0.8210 \\
                                    & ESRGAN \cite{wang2018esrgan}                      &                                    & ×4 & 28.68  & 0.8530 \\ \hline
Semi-supervised   ways              & PSSR \cite{maeda2020unpaired}                        & DIV2K \cite{agustsson2017ntire}                    & ×4 & 21.32  & 0.5541 \\\hline
\multirow{17}{*}{Unsupervised ways} & \multirow{2}{*}{DNSR \cite{zhao2018unsupervised}}      & \multirow{3}{*}{Set14 \cite{zeyde2010single}}   & ×2 & 33.83  & 0.9220 \\
                                    &                                      &                                    & ×4 & 31.76  & 0.8910 \\
                                    & DULGAN \cite{lin2018deep}                     &                                    & ×4 & 27.39  & 0.7412 \\\cline{2-6}
                                    & \multirow{2}{*}{DNSR \cite{zhao2018unsupervised}}      & \multirow{2}{*}{BSD100 \cite{martin2001database}}  & ×2 & 32.24  & 0.9010 \\
                                    &                                      &                                    & ×4 & 25.69  & 0.7880 \\\cline{2-6}
                                    & \multirow{3}{*}{CinCGAN \cite{zhang2019multiple}}   & \multirow{12}{*}{DIV2K \cite{agustsson2017ntire}}  & ×2 & 25.61  & 0.6957 \\
                                    &                                      &                                    & ×4 & 25.32  & 0.6705 \\
                                    &                                      &                                    & ×8 & 24.58  & 0.6581 \\
                                    & DNSR \cite{zhao2018unsupervised}                       &                                    & ×4 & 28.87  & 0.8650 \\
                                    & \multirow{3}{*}{MCinCGAN \cite{zhang2019multiple}}  &                                    & ×2 & 26.15  & 0.7020 \\
                                    &                                      &                                    & ×4 & 25.51  & 0.6878 \\
                                    &                                      &                                    & ×8 & 24.79  & 0.6618 \\
                                    & USISResNet \cite{prajapati2020unsupervised}                 &                                    & ×4 & 21.22  & 0.5760 \\
                                    & ULRWSR \cite{lugmayr2019unsupervised}                     &                                    & ×4 & 23.30  & 0.6200 \\
                                    & \multirow{2}{*}{KernelGAN \cite{bell2019blind}} &                                    & ×2 & 30.363 & 0.8669 \\
                                    & & & ×4 & 26.810 & 0.7316 \\
                                    & CycleSR \cite{chen2020unsupervised} & & ×4 & 23.807 & 0.5930 \\
\bottomrule
\end{tabular}}
\label{tab:15}
\end{table}

Running time and complexity are important indexes to evaluate performance of image super-resolution techniques in the real devices \cite{tian2020deep}. According to that, we conduct experiments of four GANs (i.e., ESRGAN \cite{wang2018esrgan}, PathSRGAN \cite{ma2020pathsrgan}, RankSRGAN \cite{zhang2019ranksrgan} and KernelGAN \cite{bell2019blind}) on two low-resolution images with sizes  and  for ×4 to test running time and compute parameters of different GANs. The conducted experiments have the following experimental environments. They can run on Ubuntu of 20.04.1, CPU of AMD EPYC ROME 7502P with 32 cores and Memory of 128G via PyTorch of 1.10.1 \cite{paszke2019pytorch}. Besides, they depend on a NVIDIA GeFore RTX 3090 with cuda of 11.1 and cuDNN of. 8.0.4. In Table \ref{tab:16}, we can see that ESRGAN \cite{wang2018esrgan} has slower speed than that of PathSRGAN for ×4 on image super-resolution. However, it uses less parameters than that of PathSRGAN for ×4 on image super-resolution. Thus, ESRGAN is competitive with PathSRGAN for image super-resolution. More information of different GANs for image super-resolution in terms of running time and parameters can be shown in Table \ref{tab:16}.
\begin{table}[htbp!]
\caption{Running time and parameters of different GANs for ×4.}
\label{tab:16}
\centering
\scalebox{1.0}[1.0]{
\begin{tabular}{ccc}
\toprule
Methods &Testing time (s) &Parameters\\
\midrule
ESRGAN \cite{wang2018esrgan}     & 9.7607 & 1.670×$10^{7}$ \\
PathSRGAN \cite{ma2020pathsrgan}  & 9.1608 & 2.433×$10^{7}$ \\
RankSRGAN \cite{zhang2019ranksrgan}  & 7.3818 & 1.554×$10^{6}$ \\
KernelGAN \cite{bell2019blind} & 251.00 & 1.816×$10^{5}$ \\
\bottomrule
\end{tabular}}
\end{table}

\subsubsection{Qualitative analysis of different GANs for image super-resolution}
\begin{figure}
\centering
\includegraphics[width=0.6\linewidth]{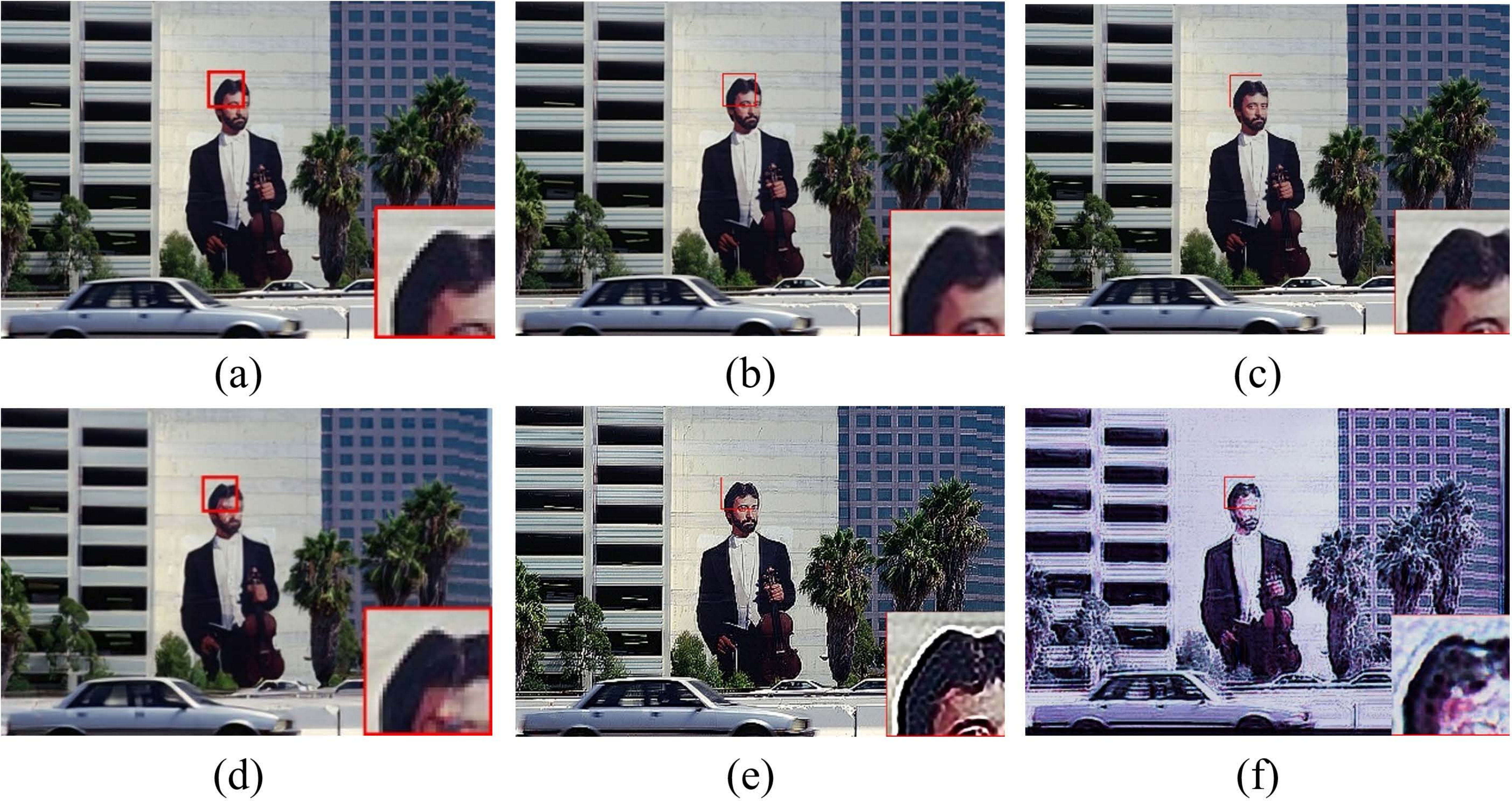}
\caption{Visual images of different GANs on an image of BSD100 for ×4: (a) original image, (b) Bicubic, (c) ESRGAN, (d) RankSRGAN, (e) KernelGAN, and (f) PathSRGAN.}
\label{fig:9}
\end{figure}

\begin{figure}
\centering
\includegraphics[width=0.6\linewidth]{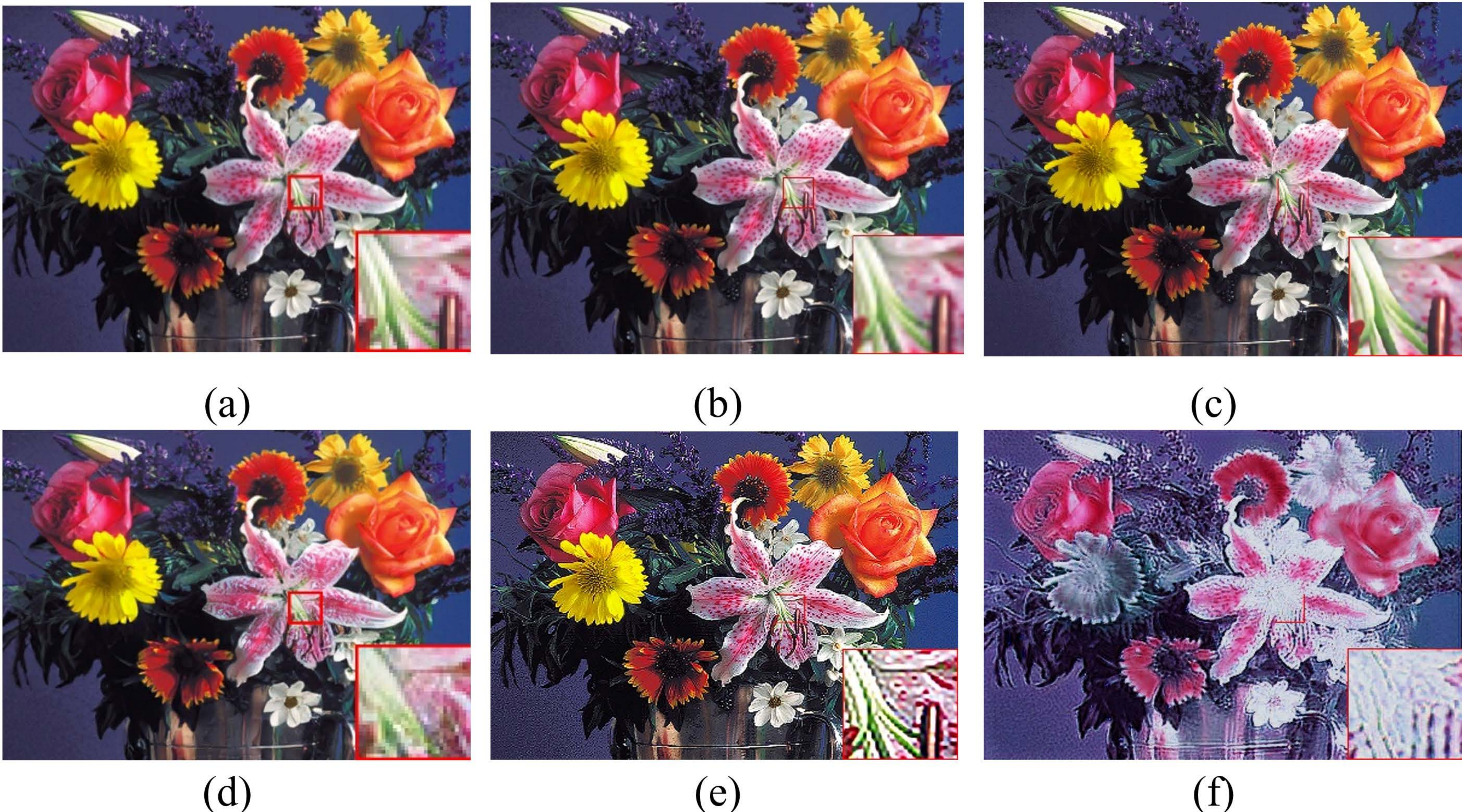}
\caption{Visual images of different GANs on an image of Set14 for ×4: (a) original image, (b) Bicubic, (c) ESRGAN, (d) RankSRGAN, (e) KernelGAN, and (f) PathSRGAN.}
\label{fig:10}
\end{figure}

To test visual effects of different GANs for image super-resolution, we choose Bicubic, ESRGAN \cite{wang2018esrgan}, RankSRGAN \cite{zhang2019ranksrgan}, KernelGAN \cite{bell2019blind} and PathSRGAN \cite{ma2020pathsrgan} to conduct experiments to obtain high-quality images for ×4. To further observe these images, we choose an area of predicted images from these GANs to amplify it as an observation area. Observation area is clearer, corresponding method has good superior SR performance. For example, ESRGAN \cite{wang2018esrgan} is clearer than that of PathSRGAN \cite{ma2020pathsrgan} on an image from the BSD100 in Fig \ref{fig:9} and Set14 in Fig \ref{fig:10} for ×4, which show that the ESRGAN is more effective in image super-resolution.

\section{Challenges and directions of GANs for image super-resolution}
Variations of GANs have achieved excellent performance in image super-resolution. Accordingly, we provide an overview of GANs for image super-resolution to offer a guide for readers to understand these methods. In this section, we analyze challenges of current GANs for image super-resolution and give corresponding solutions to facilitate the development of GANs for image super-resolution.  

Although GANs perform well in image super-resolution, they suffer from the following challenges. 

1) Unstable training. Due to the confrontation between generator and discriminator, GANs are unstable in the training process. 

2) Large computational resources and high memory consumption. A GAN is composed of a generator and discriminator, which may increase computational costs and memory consumption. This may lead to a higher demand on digital devices. 

3) High-quality images without references. Most of existing GANs relied on paired high-quality images and low-resolution images to train image super-resolution models, which may be limited by digital devices in the real world. 

4) Complex image super-resolution. Most of GANs can deal with a single task, i.e., image super-resolution and synthetic noisy image super-resolution, etc. However, collected images by digital cameras in the real world suffer from drawbacks, i.e., low-resolution and dark-lighting images, complex noisy and low-resolution images. Besides, digital cameras have higher requirement on the combination of image low-resolution and image recognition. Thus, existing GANs for image super-resolution cannot effectively repair low-resolution images of mentioned conditions.

5) Metrics of GANs for image super-resolution. Most of existing GANs used PSNR and SSIM to test super-resolution performance of GANs. However, PSNR and SSIM cannot fully measure restored images. Thus, finding effective metrics is very essential about GANs for image super-resolution. 

\textcolor{black}{6) GANs are sensitive to hyperparameters and the choice of architecture, which can affect their performance and lead to overfitting. }

\textcolor{black}{7) GANs can generate unrealistic artifacts or distortions in the output images, especially in regions with low texture or detail, which can affect the visual quality of the output images. }

\textcolor{black}{8) GAN-based image super-resolution methods may not perform well on certain types of images or degradation models.}

To address these problems, some potential research points about GANs for image super-resolution are stated below.

1) Enhancing a generator and discriminator extracts salient features to enhance     stabilities of GANs on image super-resolution. For example, using attention mechanism (i.e., Transformer \cite{han2021transformer}), residual learning operations, concatenation operations act a generator and discriminator to extract more effective features to enhance stabilities for  accelerating GAN models in image super-resolution. 
	
2) Designing lightweight GANs for image super-resolution. Reducing convolutional kernels, group convolutions, the combination of prior and shallow network architectures can decrease the complexities of GANs for image super-resolution. 
	
3) Using self-supervised methods can obtain high-quality reference images. 
	
4) Combining attributes of different low-level tasks, decomposing complex low-level tasks into a single low-level task via different stages in different GANs repairs complex low-resolution images, which can help high-level vision tasks. 
	
5) Using image quality assessment techniques as metrics evaluates quality of precited images from different GANs.

\section{Conclusion}
In this paper, we analyze and summarize GANs for image super-resolution. First, we review the development of GANs and popular GANs for image applications; then we give differences of GANs based optimization methods and discriminative learning for image super-resolution in terms of supervised, semi-supervised and unsupervised manners. Next, we compare the performance of these popular GANs on public datasets via quantitative and qualitative analysis in SISR. Then, we highlight challenges of GANs and potential research points on SISR. Finally, we summarize the whole paper.

\begin{acks}
This work was supported in part by the National Natural Science Foundation of China under Grants 62576123, in part by CAAI-CANN Open Fund developed on OpenI Community, and in part by the Natural Science Foundation of Heilongjiang Province under Grant YQ2025F003.
\end{acks}

\bibliographystyle{unsrt}
\bibliography{sample-acmsmall}
\end{document}